\documentclass[12pt,preprint]{aastex}

\slugcomment{Accepted for publication in The Astrophysical Journal}

\shorttitle{The evolved pulsating CEMP star HD\,112869}
\shortauthors{Za\v{c}s et al.}

\begin{document}

\title{The evolved pulsating CEMP star HD\,112869}

\author{Laimons Za\v{c}s\altaffilmark{1,2}, Julius Sperauskas\altaffilmark{1},
        Aija Grankina\altaffilmark{2},  Viktoras Deveikis\altaffilmark{1},
        Bogdan Kaminskyi\altaffilmark{3},  Yakiv Pavlenko\altaffilmark{3}, 
        and Faig A.Musaev\altaffilmark{4,5,6}}

\altaffiltext{1}{Vilnius University Observatory, \v{C}iurlionio 29, Vilnius 2009, Lithuania}
\altaffiltext{2}{Laser Center, University of Latvia, Rai\c{n}a bulv\= aris 19, LV-1586~R\=\i ga,Latvia }
\altaffiltext{3}{Main Astronomical Observatory of Academy of Sciences of Ukraine,  Zabolotnoho 27, Kyiv, 03680, Ukraine }
\altaffiltext{4}{Special Astrophysical Observatory of the Russian AS, Nizhnij Arkhyz, 369167, Russia }
\altaffiltext{5}{ Institute of Astronomy of the Russian AS, 48 Pyatnitskaya st., 
           119017, Moscow, Russia }
\altaffiltext{6}{ Terskol Branch of Institute of Astronomy of the Russian AS,
           361605 Peak Terskol, Kabardino-Balkaria, Russia }

\begin{abstract}
Radial velocity measurements,  $BVR_C$ photometry, and high-resolution spectroscopy in the wavelength region  from blue to near infrared are employed in order to clarify the evolutionary status of the carbon-enhanced metal-poor star HD\,112869 with unique ratio of carbon isotopes in the atmosphere.  An LTE abundance analysis was carried out using the method of spectral synthesis and new self consistent 1D atmospheric models. The radial velocity monitoring confirmed semiregular variations with a peak-to-peak amplitude of about 10 km\,$s^{-1}$ and a dominating period of about 115 days. The light, color and radial velocity variations are typical of the evolved pulsating stars. The atmosphere of HD\,112869 appears to be less metal-poor than reported before,  [Fe/H] = $-$2.3 $\pm$0.2 dex. Carbon to oxygen and carbon isotope ratios are found to be extremely high, C/O $\simeq$ 12.6 and $^{12}C/^{13}C  \gtrsim$ 1500, respectively. The s--process elements yttrium and barium are not enhanced, but neodymium appears to be overabundant.  The magnesium abundance seems to be lower than the average found for CEMP stars, [Mg/Fe] $<$ +0.4 dex. HD\,112869 could be a single low mass halo star in the stage of asymptotic giant branch evolution.
\end{abstract}

\keywords{stars: carbon -- stars: AGB and post-AGB -- stars: abundances --  stars: oscillations --
                  stars: individual: TT\,CVn}

\section{Introduction}

The wide-field spectroscopic surveys revealed high frequency of carbon-enhanced stars with [C/Fe] $>$ 1.0 (hereafter CEMP stars) among metal-poor stars \citep{beers92, christlieb01}. According to \citet{lucatello05, lucatello06}, up to 30 percents of stars with [Fe/H] $<$ --2.5 are carbon-rich. The chemical abundances of CEMP stars reflect the origin of CNO and neutron capture elements  in the early Galaxy. CEMP stars have a wide variety of elemental abundance patterns {\bf  (see \citet{beers07}; and references therein)}. The majority of CEMP stars have enhanced s-process elements \citep{aoki07} and are referred to as CEMP-s stars. Other CEMP stars exhibit strong enhancements of r-process elements (CEMP-r) or the presence of enhanced neutron-capture elements associated with both the r- and s$-$processes (CEMP-r/s). The class of CEMP-no stars comprises stars that, in spite of high carbon (and often N and O) overabundances with respect to Fe, do not exhibit strong neutron-capture elements. Recently, CEMP stars have also been found with large enhancements of $\alpha$-elements \citep{norris01, aoki02, depagne02}, which \citet{aoki06} refer to as CEMP-$\alpha$ stars.

Observed CEMP stars bear signatures of thermally pulsating asymptotic giant branch (TP-AGB) and explosive nucleosynthesis in the early Galaxy and subsequent mixing or mass transfer in single or binary stars (see \citet{masseron10}; and references therein). The carbon in the CEMP-s stars is very likely to have been produced by an intermediate-mass asymptotic giant branch (AGB) companion that has transferred material to the currently observed star. \citet{bisterzo} suggest that all CEMP-s stars observed until now are old halo main sequence and turn-off stars or giants of a low initial mass (M $<  0.9M_{\sun}$).
The origin of the carbon in the other CEMP classes is not yet fully understood. The CEMP-no stars are of special interest because  the observed CNO elements probably have been produced by a primordial population of massive (20 $<  M/M_{\sun} <$ 100) and megametal-poor ([Fe/H] $<$ --6) stars, which are predicted to have experienced significant mass loss of CNO-enhanced material  via strong winds \citep{hirschi, meynet}. Alternatively, the carbon excess at extremely low metallicity may have been produced by the ejecta of the so-called faint supernovae \citep{umeda}.

A high $^{12}C/^{13}C$ ratio is observed rarely among the analyzed CEMP stars. The CEMP-no stars exhibit quite low $^{12}C/^{13}C$ ratios (in the range 4 $<$ $^{12}C/^{13}C$ $<$ 10), indicating that a significant amount of mixing has occurred in their progenitor objects \citep{aoki07, sivarani}. Strong mixing occurs on the red giant branch due to the first dredge-up (FDU) affecting carbon and nitrogen abundances and decreasing the $^{12}C/^{13}C$ ratio. However, it is not sufficient to explain the low isotopic ratios $^{12}C/^{13}C <$ 30 observed for most of the studied CEMP-s and CEMP-r/s  stars (see \citet{bisterzo}; and references therein).  Further extra-mixing has to be hypothesized during the following AGB phase to interpret the observations. Cool bottom processing (CBP) may explain the observed [C/Fe] and [N/Fe]. According to \citet{bisterzo}, most of the studied CEMP-s and CEMP-r/s stars need the occurrence of CBP to describe the observed abundance patterns.  The observed [C/Fe] ratios agree with AGB predictions without involving of extra-mixing only in four stars: CS22880$-$074, CS30315$-$91, HE0507$-$1653, and HE1429$-$0551. The first two stars  show, in addition, relatively high lower limits for $^{12}C/^{13}C$ ($>$40--60). HD\,13826 (= V\,Ari) is an exception, with high $^{12}C/^{13}C$ ratio found by \citet{kipper90} and confirmed by \citet{aoki} and \citet{beers07}, $^{12}C/^{13}C$ = 90 $\pm$ 10. \citet{tsuji} and \citet{kipper} detected another metal poor star, HD\,112869 (= TT\,CVn),  with an extremely low intensity of molecular lines with isotopic $^{13}C$. 

HD\,112869  is a high latitude (b = +79$\degr$) extremely metal-poor  ([Fe/H] = $-$2.9; Kipper \citet{kipper}) star classified as R type carbon star with enhanced CH bands by \citet{keenan}.  The bolometric magnitude estimated by \citet{bergeat02}, $M_{bol}$ = --3.35 mag, is in a good agreement with an independent estimation provided by \citet{ eggen}, $M_{bol}$ = --3.5 mag. Estimation of the trigonometric distance is not possible because  of a negative parallax value for HD\,112869, $\pi = -1.26 \pm$ 1.28 mas \citep{vanLeeuwen}. Effective temperature for HD\,112869 was determined by means of the infrared flux method (IRFM), which uses the ratio of bolometric flux to the  infrared flux in L band, $T_\mathrm{eff} = 3715~\mathrm{K}$ \citep{aoki}. The effective temperature found by \citet{bergeat01} using the same method is a bit higher,  $T_{eff}$ = 3870 K.  \citet{tanaka} calculated the effective temperature, $T_{eff}$ = 4100 K,  by fitting the near infrared spectra at wavelengths of 1.35, 1.74, and 2.29 $\mu$m, where the molecular absorption because of CO, CN, and $C_2$ is relatively low, with the model spectra. Surface gravity, $\log g$ = 0.4 (cgs), was estimated by \citet{kipper} using the well known relation between  $\log g$, $T_{eff}$, stellar mass (M), and $M_{bol}$, assuming for the star 1$M_{\sun}$.  \citet{aoki} used for spectral synthesis two different values of $\log g$,  0.0 and 2.0 (cgs). The microturbulent velocity ($\xi_t$) was estimated iteratively using the profiles of molecular lines \citep{kipper} and theoretical curves of growth for CN lines \citep{aoki} in the range from 3 to 6 km $s^{-1}$. Abundances for HD\,112869 have been calculated using methods of high-resolution spectroscopy by \citet{ tsuji, kipper, aoki}, however, the results are contradictory (see discussion in Section 3.3.12). HD\,112869 exhibits semiregular light variations (SRB) with an amplitude of about 0.5 mag in the visual band \citep{samus} and a period of about 105 days was suspected. Five radial velocities measurements by R.F.Griffin revealed velocity variations in the range from --132.6 to --136.9  km\,$s^{-1}$ \citep{yoss}. Mass transfer from an AGB companion star in a binary system \citep{kipper} or evolution of a single star on the AGB \citep{tsuji, aoki} are proposed to explain the observed properties of HD\,112869. The basic data for HD\,112869 and the comparison stars, available in the literature, are summarized in Table~\ref{basic_dat}.

In this paper the results of contemporaneous radial velocity monitoring and broadband $BVR_C$ photometry for HD\,112869 are presented. The  observations are analyzed to clarify the character of variability and to search for its period. New high resolution optical spectrum was observed  and analyzed using qualitative and quantitative methods relative to the comparison stars:  well known N-type carbon star HD\,92055 and J -type carbon star HD\,25408 with enhanced $^{13}C$ isotopic abundance. Abundances were calculated for carbon, nitrogen, magnesium, iron, and  selected neutron-capture elements, using the method of spectral synthesis and new self-consistent atmospheric models. The evolutionary status of HD\,112869 is discussed.

\begin{table*}
\begin{center}
\caption{The basic data for HD\,112869 and spectroscopic comparison stars collected from literature.\label{basic_dat}} 
\begin{tabular}{cccccccccccc}
\tableline\tableline
 HD &  Name & Sp.Type  & $M_{bol}$  & $T_{eff}$ & [Fe/H] & C/O & $^{12}C/^{13}C$ & [s/Fe]\tablenotemark{a}  & [hs/ls]\tablenotemark{b}  & Ref \\
\tableline
112869  & TT\,CVn & C\,4,5 CH  &  -3.35  & 3700  &  -2.9   & 1.07  &  50-90  & +2.5  &  +0.8  & 1   \\
              &               &                   &                  &  3700 & $\eqsim$ -2.0 & 6.3   & $\geqslant$500 &  ...       &     ...    &  2  \\
25408 &  UV\,Cam  & C\,5,3 CH & -4.85 &  3000 & -0.82 & 1.20 &  5  &  +1.2 & +0.7 &   3 \\
      &         &        &      -8.70 &    3350    &    +0.2\tablenotemark{c} &  ...     &    4     &   +0.09\tablenotemark{d} & ...  &   4  \\
92055  & U\,Hya  & C\,6,3 & -4.0 &  2825 & -0.05 & 1.05  & 35   &  +1.0  & 0.0  &  5  \\
\tableline
\end{tabular}
\tablenotetext{a}{s = $<$ls + hs $>$}
\tablenotetext{b}{hs = $<$Ba, La, Ce, Nd, Sm$>$,  ls = $<$Sr, Y, Zr$>$}
\tablenotetext{c}{mean metallicity: [M/H]}
\tablenotetext{d}{[s/M]}
\tablerefs{(1)~ \citet{kipper}; (2)~ \citet{aoki}; (3)~\citet{ kipper96}; (4)~ \citet{abia00}; (5)~\citet{abia02}}
\end{center}
\end{table*}

\section{Observations and data reduction}

Radial-velocity monitoring of HD\,112869 was started in 2006 using the CORAVEL spectrometer of the Vilnius University installed on the 
1.65-m telescope at the Mol{\.e}tai Observatory (Lithuania). The CORAVEL spectrometer \citep{upgren} is based on 
principles of the photoelectric radial-velocity scanner developed by R.F.Griffin \citep{griffin} and it operates by scanning the spectrum of a star across
the mask and obtaining an on-line cross-correlation velocity. 97 measurements are gathered for HD\,112869 during about 8 years of monitoring. The standard deviation 
of a single observation for late-type stars brighter than about 11th magnitude is usually better than 0.8~km s$^{-1}$.  The velocities 
have been standardized using the observations of IAU radial velocity standards \citep{udry1} and are close to the system of velocities published by \citet{nidever} and \citet{marcy}.  The difference in zero point was found to be 0.14~km s$^{-1}$ for 
F-G-K type stars and 0.4~km s$^{-1}$ for M-type stars and exhibits rms scatters of 0.5 and 0.8 kms$^{-1}$, respectively. 

Photometric $BVR_{\rm C}$ observations were performed from  2013 February to 2014 June at the Mol{\.e}tai Observatory using the twin-telescope photometric system. It consists of the 63-cm and 25-cm telescopes for simultaneous measurements of a target star and of a comparison star, respectively, and is equipped with two identical photoelectric (multi-alkali
photocathodes of type S20) photometers connected to the same data acquisition system. The ratio of counting efficiency of one
telescope to that of another was checked from time to time by measuring the same star. The comparison stars, HD\,114036 and
HD\,114357, are chosen to lie within 2$\degr$ from HD\,112869  and are bright enough to be accessible with 
the 25-cm telescope. The simultaneous measurements and the proximity of the comparison stars to HD\,112869  reduced the effects of variations of atmospheric transparency.

A set of 12 standard stars of spectral types from late F to K5 was observed repeatedly at various air masses to measure atmospheric extinction and to place the data of differential photometry on a standard magnitude scale. Both the extinction coefficients
and the transformation to the standard Johnson-Cousins system were then calculated using a least-squares routine which solves
simultaneously all repeated observations. Since the primary standards in the Cousins system \citep{landolt92} are not accessible
from the Mol{\.e}tai Observatory sky, for the transformation of instrumental $R_C$ magnitude we addressed the Catalog of $WBVR$
magnitudes of bright stars of the northern sky by \citet{WBVR91},
in which all of these 12 standard stars are found. The $R$ magnitude of the $WBVR$ catalog, however, is not in the Cousins system,
therefore we transformed it to $R_{\rm C}$ using an additional set of 19 stars having both the Cousins and $WBVR$ photometry. The uncertainty of the transformation equation obtained for R is 0.018 mag. Standardized $BVR_{\rm C}$  photometry  of the comparison stars is listed in Table~\ref{comp_stars}. The bottom line of the table gives an overall error in each magnitude, estimated taking into account both the measurement error and the transformation uncertainties. These stars should serve as useful comparison objects for photometric monitoring of HD\,112869 in the future.

\begin{table}
\begin{center}
\caption[]{Estimated magnitudes and overall errors for the photometric comparison stars.\label{comp_stars}}
\begin{tabular}{ccccc}
\tableline\tableline
Star & Sp.Type  & B & V & $R_C$ \\
        &                & (mag) & (mag) & (mag) \\
\tableline
HD\,114036  & G8\,V    & 9.010  & 8.153  & 7.587 \\
HD\,114357  &  K2\,III  & 7.247  & 6.014  & 5.270 \\
Error             &               & 0.018  & 0.015  & 0.024 \\
\tableline
\end{tabular}
\end{center}
\end{table}

\begin{figure}  
\resizebox{\hsize}{!}{\includegraphics{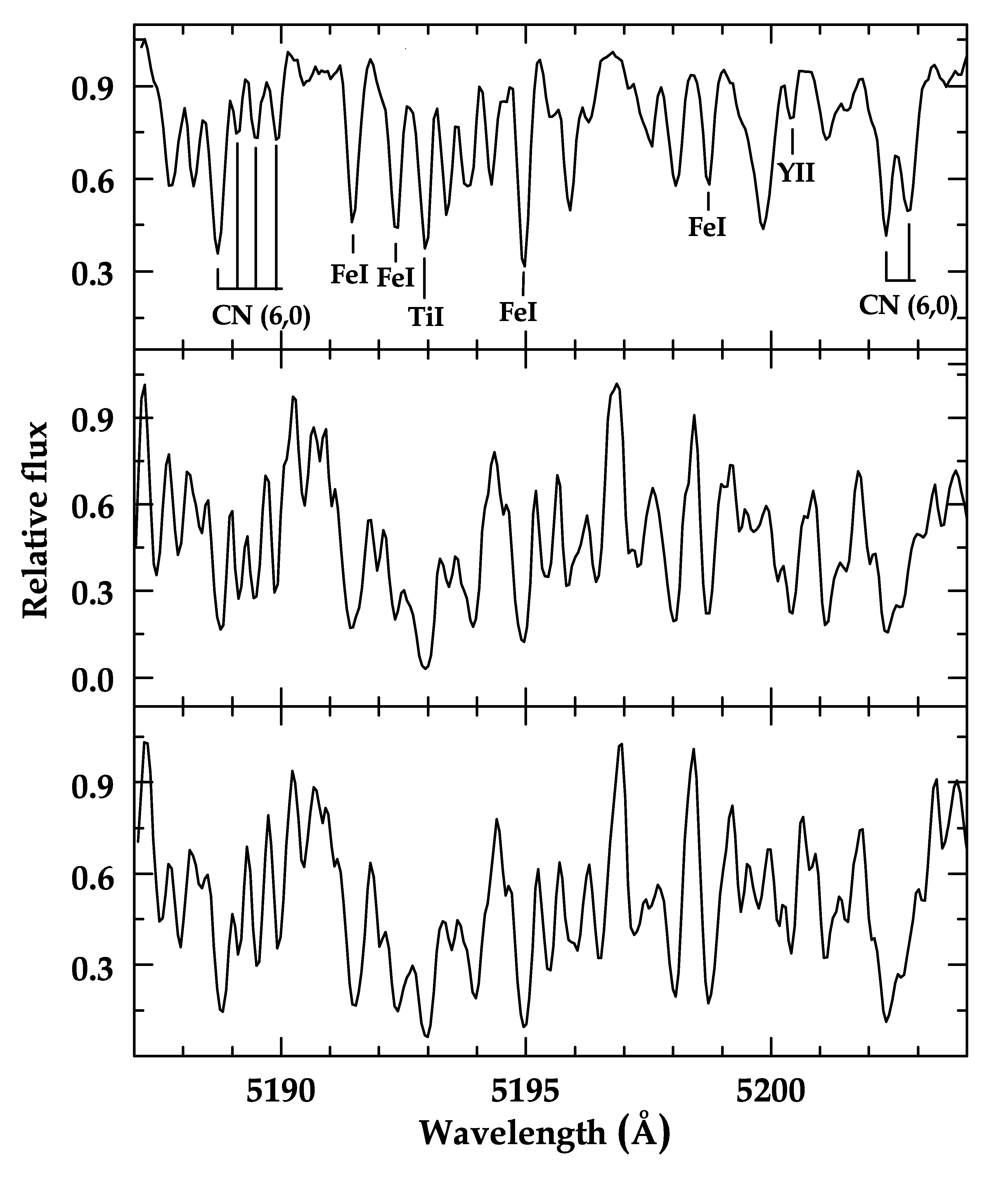}}
\caption{Spectrum of HD\,112869 (top panel), along with that for the comparison stars HD\,92055 (middle panel) and HD\,25408 (bottom panel), in the spectral region less blended by molecular features.  Iron and yttrium lines are relatively weak in the spectrum of HD\,112869 in comparison with those for the comparison stars.}
\label{fig1}
\end{figure}

\begin{figure}  
\resizebox{\hsize}{!}{\includegraphics{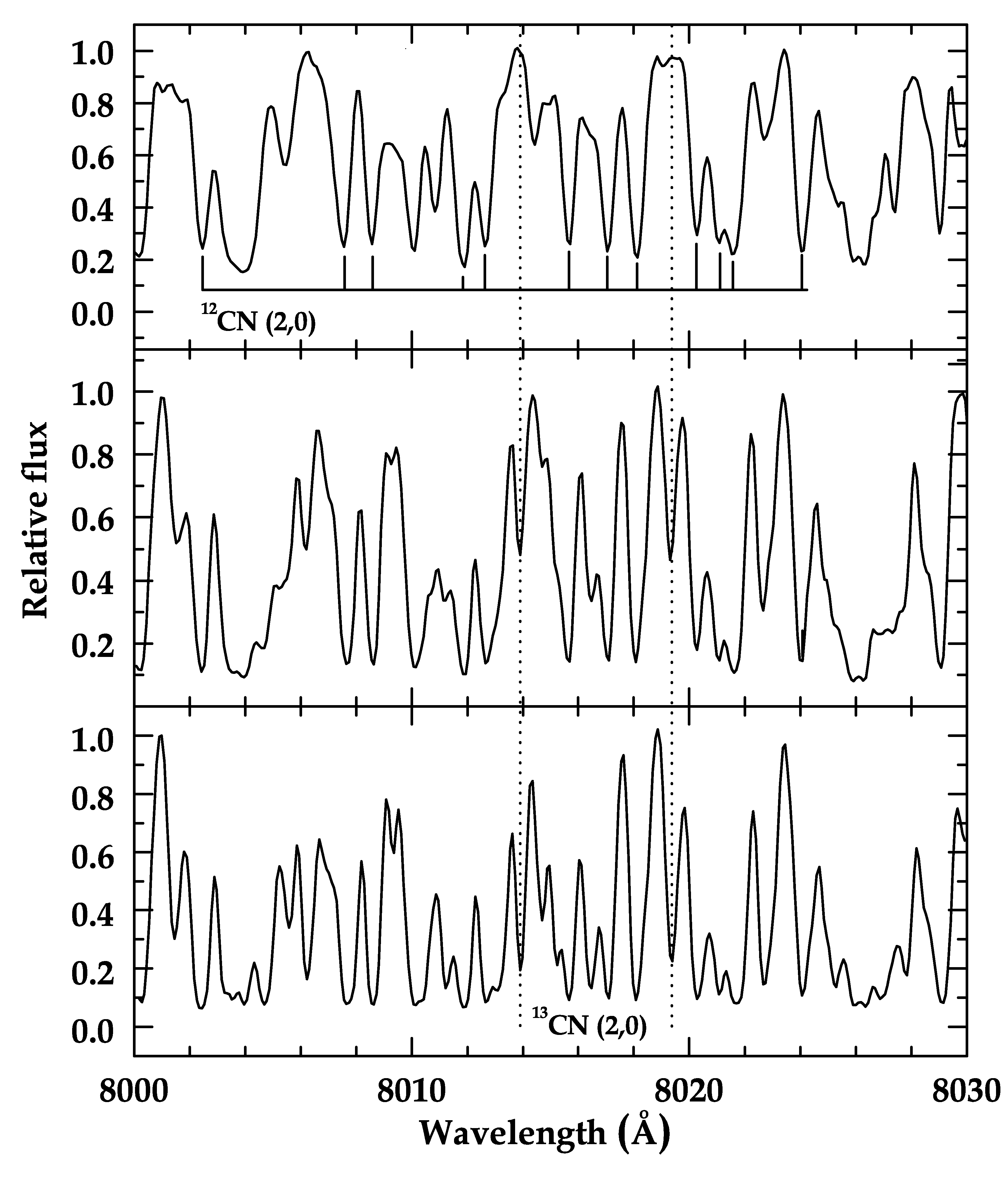}}                                      
\caption{Same as Fig.~\ref{fig1}, but in the region of selected CN(2,0) lines. The isotopic $^{13}$CN lines are not clearly visible in the spectrum of HD\,112869. Thus a high $^{12}$C/$^{13}$C ratio is confirmed in comparison with that of  HD\,92055 ($^{12}$C/$^{13}$C = 35) and HD\,25408 ($^{12}$C/$^{13}$C = 4).}
\label{fig2}
\end{figure}

The high-resolution spectra for HD\,112869 and for the spectroscopic comparison stars HD\,92055 and HD\,25408 were observed with the coud\'e \'echelle spectrometer MAESTRO fed by the 2-m telescope at the Observatory on the Terskol Peak in Northern Caucasus, equipped with a CCD detector and having a resolving power of $\sim$45\,000. The spectrum of HD\,112869 was observed during an exposure of 7200 s with S/N ratio of about 80  near $H_{\alpha}$ on the night of 2013 March 8. The spectra of both comparison stars are retriewed from the MAESTRO archive of high resolution spectra. These observations were carried out in 2001 January and 2003 February, respectively. All the employed spectra cover the wavelength region from about 3600 to 9300 \AA\  in 85 wavelength bands overlapping in the red and near infrared wavelengths regions. All the spectra were bias subtracted, flat-field corrected, and converted to one-dimensional spectra using the standard DECH20T package.\footnote{http://www.gazinur.com/Spectra-Processing.html}  The wavelength calibration was made using Th$-$Ar and sky spectra obtained for each night. In addition, a spectrum of a hot and rapidly rotating star was observed to identify the telluric absorption lines.  Two representative wavelength regions for HD\,112869, along with those for the comparison stars, are shown in Figures~\ref{fig1} and \ref{fig2}.

\section{Analysis and results}

\subsection{Radial velocities}

\begin{figure*}  
 \resizebox{\hsize}{!}{\includegraphics{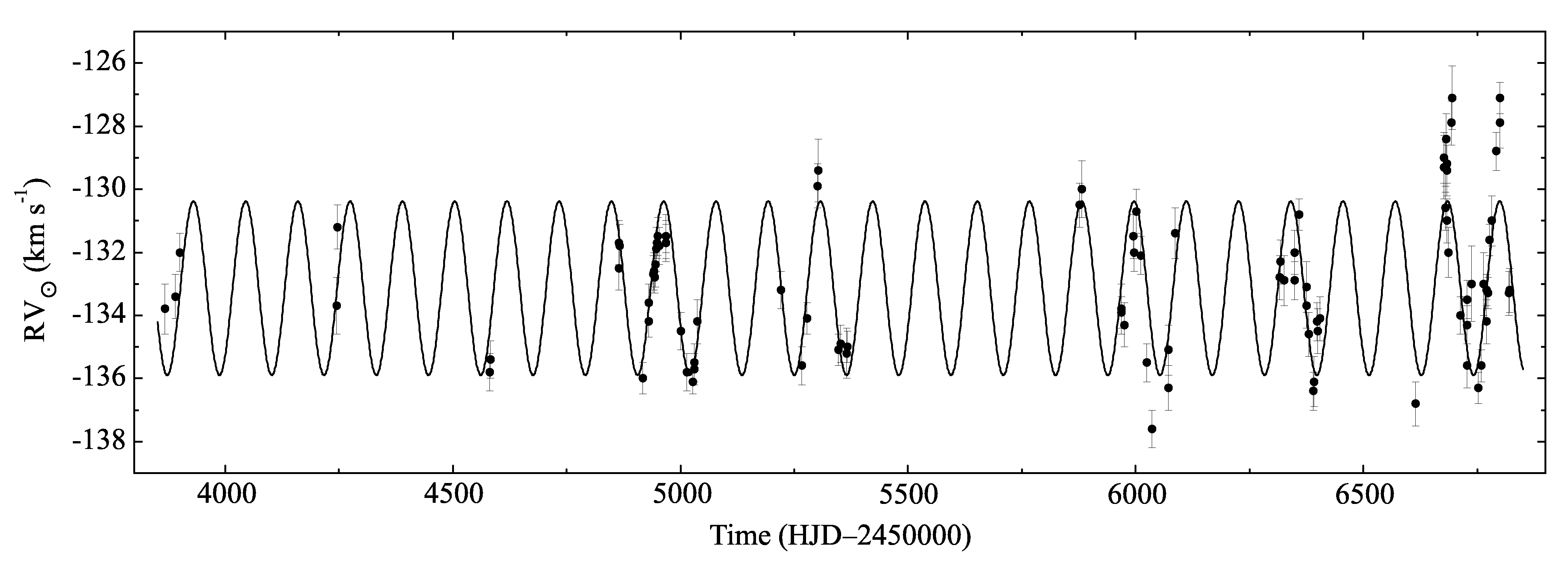}}
                              \caption{Heliocentric radial velocity of HD\,112869 measured using the CORAVEL spectrometer (filled circles),  illustrating a semiregular character of variability. The observations are approximed by a sine wave with  a period of  P = 114.9 days and an amplitude of A = 2.8~km s$^{-1}$.}
  \label{fig3}
\end{figure*}

\begin{figure}
\resizebox{\hsize}{!}{\includegraphics{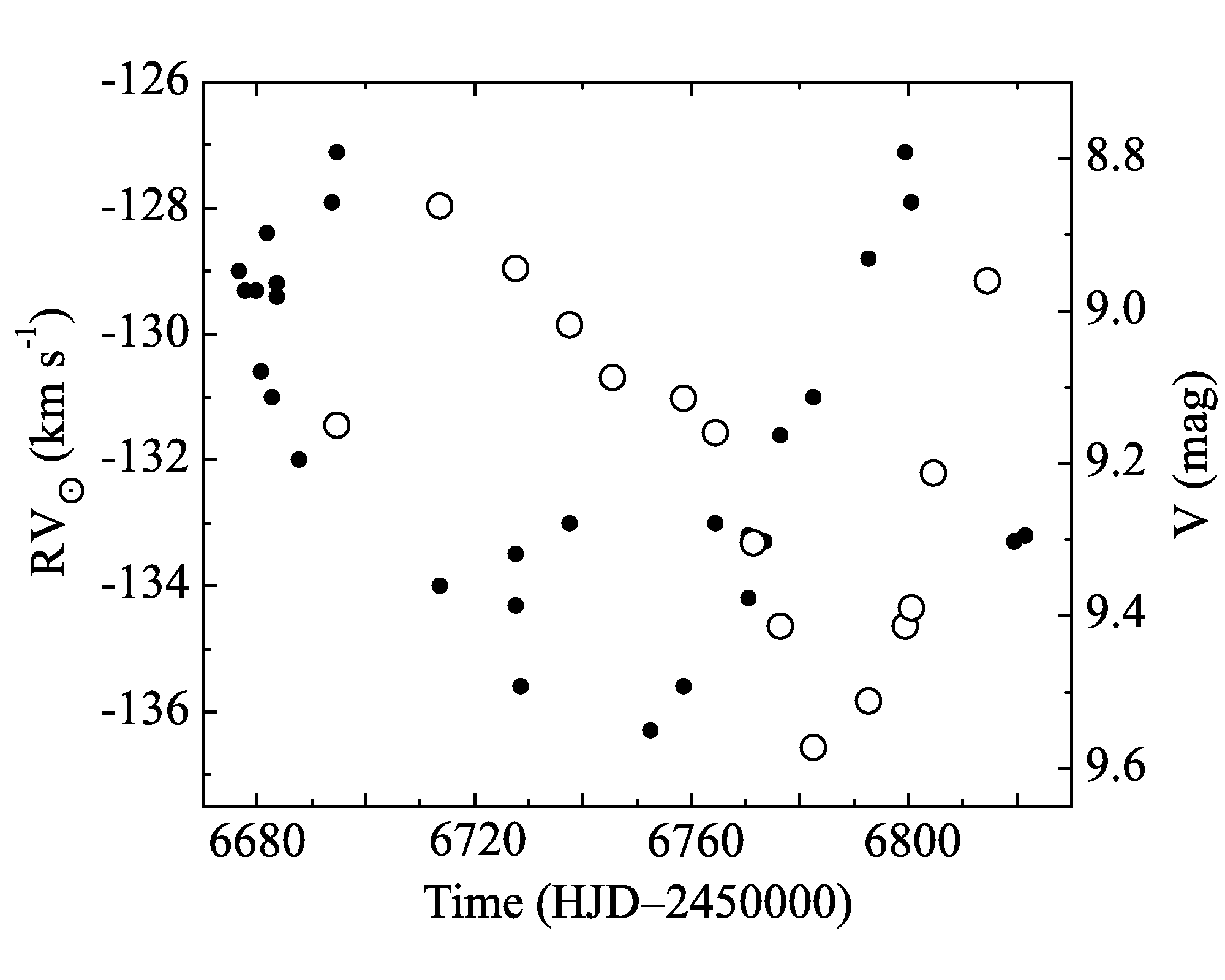}}
    \caption{Heliocentric radial velocity of HD\,112869 measured using the  CORAVEL spectrometer (filled circles)  along with
the apparent magnitude in V band (open circles) for the last observed cycle, confirming the pulsational nature and semiregular character of the variability.} 
     \label{fig4}
\end{figure}

\begin{figure}
\resizebox{\hsize}{20cm}{\includegraphics{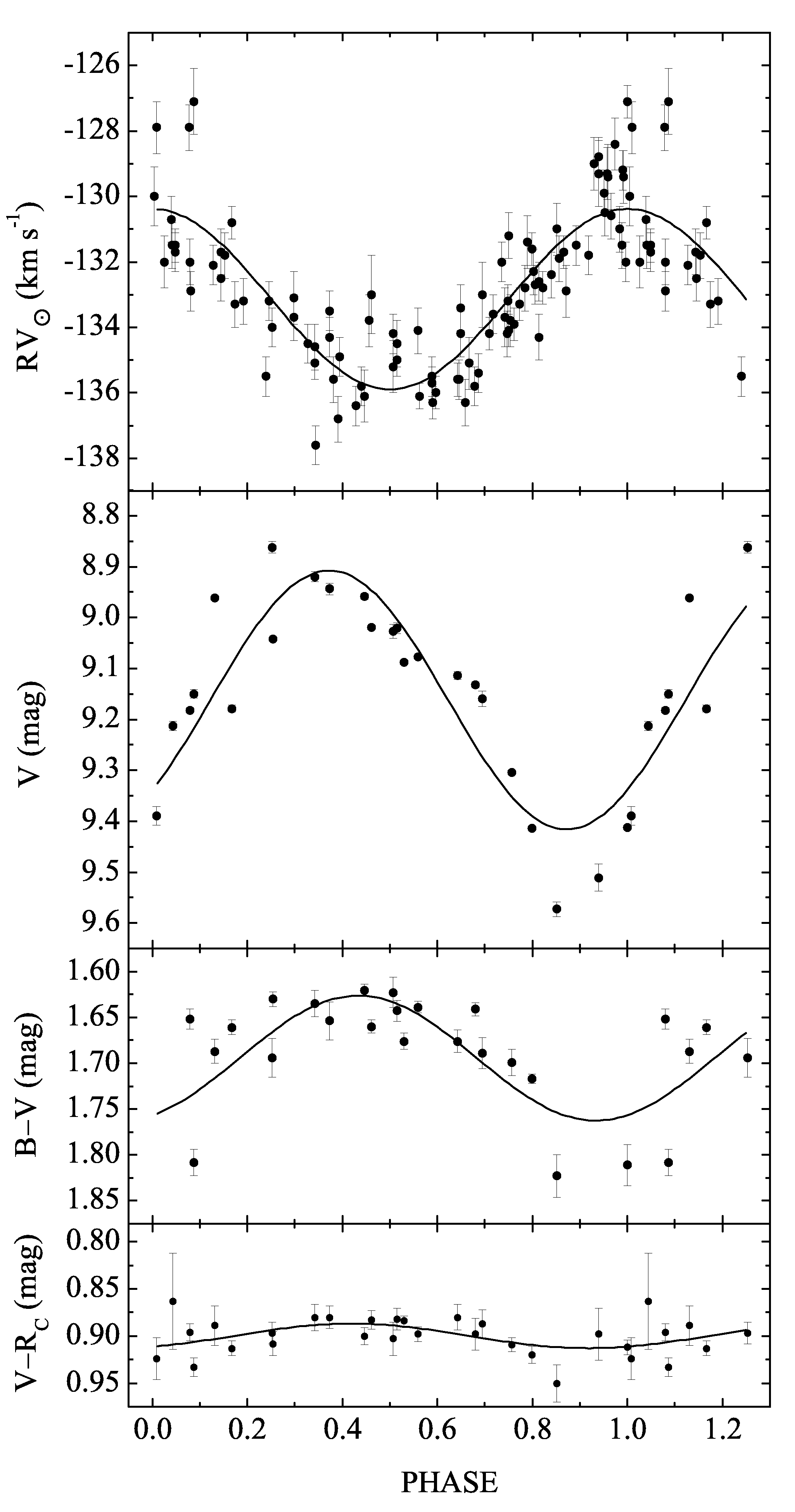}}
    \caption{CORAVEL radial velocities  (top panel) and photometry for HD\,112869 phased with a period of 114.9 days. 
                  The fit with a sine wave is shown.} 
     \label{fig5}
\end{figure}

Heliocentric radial velocities ($RV_{\sun}$) for HD\,112869  as a function of Heliocentric Julian Date (HJD), measured using the CORAVEL spectrometer, are listed in Table~\ref{RV}, along with the uncertainty of each observation. The radial velocity curve is shown in Figure~\ref{fig3}. As can be seen, a variation in the radial velocities reaches about 10 km s$^{-1}$. 
A periodogram analysis was carried out of the radial velocity data set of 97 data points and a significant period was detected, P = 114.9 days. However, the character of
variability varies from cycle to cycle. During the  time span HJD 2454850-5050 the variations are quite regular, however, recent radial velocity data display a significant deviation from a sine wave  (see Figures~\ref{fig3} and ~\ref{fig4}). It can be seen also that the scatter of velocities  near the velocity extremums sometimes exceeds remarkably the error of measurements and such a large scatter probably reflects some physical processes in the atmosphere/envelope of the star. All the velocity measurements for HD\,112869, as a function of phase, resulting from the approximation by a sine wave with a period of 114.9 days, are plotted in Figure~\ref{fig5} (top panel).  The radial-velocity monitoring shows with certainty that HD\,112869 is a pulsator with a dominating pulsation period of about 115 days and a variability pattern typical of the evolved low mass stars (see \citet{hrivnak} and references therein).

\begin{table*}
\begin{center}
\caption[]{CORAVEL radial-velocity measurements for HD\,112869 along with the uncertainties.\label{RV}}
\begin{tabular}{ccccccccc} 
\tableline\tableline
HJD & $RV_{\sun}$  & $\sigma$  & HJD  & $RV_{\sun}$ &  $\sigma$  &  HJD  & $RV_{\sun}$ &  $\sigma$ \\
-2,450,000 & (km s$^{-1}$)     &  (km s$^{-1}$)  & &  &  &  &  &        \\
\tableline
3867.361  & -133.8  &    0.9 & 5301.544  & -130.1  & 0.7  & 6676.590 &  -129.0   &  0.8  \\
3889.428  & -132.7  &    0.8 & 5302.50   & -129.8   & 0.8 & 6677.617  & -129.3    & 1.0  \\
3899.386  & -132.0  &    0.8 & 5346.366 &  -135.5  & 0.5  & 6679.660 &  -129.3   &  0.8  \\
4244.423  & -132.9  &    1.1 & 5352.346 & -135.4   & 0.6 & 6680.646 &  -130.6    &  0.7 \\
4245.397  & -132.0  &    0.7 & 5365.371 & -135.4   & 0.8 & 6681.639 & -128.4     & 0.8  \\
4581.453  & -136.4  &    0.5 & 5366.350 &  -135.4 & 0.5  &  6682.686 & -131.0    & 0.7  \\
4582.417  & -135.8  &    0.5 & 5875.691 & -130.9  & 0.7  & 6683.582  & -129.2    & 0.6 \\
4864.573  & -132.1  &    0.7 & 5881.665 & -130.6  & 0.9  & 6683.609  &   -129.4  & 0.8   \\
4864.591  & -132.5  &    0.6 &  5967.573 & -134.2  & 0.7  & 6687.586 &  -132.0    &  0.8 \\
4865.582  & -132.2  &    0.7 & 5968.611 & -134.3   & 0.5  & 6693.594 &  -127.9    &  0.7 \\
4916.475  & -135.9  &    0.5 & 5974.617 & -134.4   & 0.7  & 6694.600 &  -127.1    &  1.0 \\
4929.381  & -134.7  &    0.5 & 5994.551 & -131.2  & 0.7   & 6713.533 &  -134.0    &  0.6 \\
4930.295  & -133.9  &    0.6 & 5995.473 & -131.7   & 0.6  & 6727.511 &  -133.5    &  0.6 \\
4940.435  & -132.8  &    0.6 & 6000.476  &  -130.4  & 0.7 & 6727.523 &  -134.3    &  0.6 \\
4941.472  & -132.4  &    0.5 & 6010.550  & -131.8  & 0.6  & 6728.483 &  -135.6    &  0.7 \\
4942.413  & -133.0  &    0.5 & 6023.407  &-135.2   & 0.6  & 6737.497  & -133.0    &  1.2 \\
4944.348  & -132.6  &    0.7 & 6035.422  & -137.3   & 0.6  & 6737.572  & -134.8    &  1.5 \\
4946.393  & -132.1  &    0.6 &  6071.427 & -136.0   & 0.7  & 6752.453 &  -136.3    &  0.5 \\
4947.389  & -132.0  &    0.5 & 6072.375  & -134.8   &  0.8  & 6758.379 &  -136.4    &  0.5 \\
4950.401  & -131.8  &    0.5 & 6086.444  & -131.1   &  0.8  & 6758.409 &  -135.6    &  0.5 \\ 
4953.422  & -131.8  &    0.6 & 6315.595  &  -132.5  &  0.7  & 6764.396 &  -133.8    &  0.6 \\
4967.516  & -131.6  &    0.7 & 6317.569  &   -132.0  &  0.7  & 6770.372  & -134.2    &  0.7 \\
4968.325  & -131.3  &    0.6 & 6325.433  & -132.6   &  0.8   & 6770.468  & -133.2    &  0.5 \\
4968.332  & -131.5  &    0.6 & 6349.447  & -131.7   &  0.7   & 6773.384  & -133.3    &  0.5 \\
5000.378  & -134.6  &    0.6 &  6359.414 & -130.5   &  0.5   & 6776.394  & -131.6    &  0.5 \\
5013.371  & -136.0  &    0.6 & 6374.471  &  -133.4  &  0.7   & 6782.360  & -131.0    &  0.8 \\
5027.345  & -136.2  &    0.5 & 6379.454  & -134.3   &  0.7   & 6792.431 &  -128.8   &   0.6 \\
5030.348  & -135.7  &    0.5 & 6389.552  & -136.1   &  0.6  & 6799.344 &  -127.1    &  0.5 \\
5030.368  & -135.5  &    0.6 & 6391.497  & -135.8   &  0.8  & 6800.511  & -127.9    &  0.8 \\
5037.384  & -134.0  &    0.7 & 6398.373  &  -133.9  &  0.6   & 6819.376  & -133.3    &  0.7 \\
5220.668  & -133.4  &    0.6 & 6399.435  &  -134.2  &  0.7  & 6821.370  & -133.2    &  0.7  \\
5266.505  & -136.3  &    0.6 &  6404.398 &  -133.8  &  0.7  & ...             &   ...         &    ...   \\    
5278.481  & -134.3  &    0.5 &  6614.668 &  -136.8 &  0.7  &     ...         &    ...         &  ...    \\     
\tableline
\end{tabular}
\end{center}
\end{table*}

\subsection{Photometry}

The results of BV$R_C$ photometry are given in Table~\ref{phot}. The columns contain the following information: time of the measurement, the mean magnitude V and color indices (B-V) and  (V-$R_C$) averaged over each night, and the corresponding standard errors. It should be noted that the errors given are photometric uncertainties and reflect primarily the actual observing conditions. The light curves show a cyclical variations with amplitudes of $\sim$0.84 mag in B, $\sim$0.71 mag in V, and $\sim$0.66 mag in $R_C$. The light variations in V band are shown in Figure~\ref{fig4}, along with the velocity measurements during the last observed cycle of pulsations.  A large scatter in the velocities ($\sim 6\sigma$) near HJD 2456680 and a significant departure from a sine wave in the V light are clearly seen. All the photometric data phased with a  period  of 114.9 days are given in Figure~\ref{fig5}. As it can be seen, the dominating period of pulsations calculated using photometry is close to that found on the basis of radial velocities. The  light variations are accompanied by color variations what lends observational support to the stellar pulsations (Figures~\ref{fig4} and \ref{fig5}). The color curves mimic the cyclical variations in the light curve, with the star seen to be redder when it becomes fainter.  A comparison of the light, color, and velocity curves reveals that the light and color curves are about in phase, while the radial velocity curve is out of phase with the light curve. Thus, the radial velocity and light/color variations display the correlations observed for the pulsating evolved stars (see \citet{hrivnak}; and references therein).

\begin{table*}
\begin{center}
\caption{BV$R_C$ photometry for HD\,112869 along with the standard deviations.\label{phot}} 
\begin{tabular}{ccccccc}
\tableline\tableline 
HJD & V & $\sigma_V$ & $B-V$ & $\sigma_{(B-V)}$ &  $V-R_C$ & $\sigma_{(V-R_C)}$ \\
$-$2,450,000 & (mag) & (mag) &  (mag) & (mag) &  (mag) & (mag)   \\
\tableline
6349.427 &  9.182  & 0.005 &  1.652 & 0.011 &  0.896  &  0.009 \\
6359.439 &  9.179  & 0.007 &  1.661 & 0.008 &  0.913 &  0.008 \\
6369.445 &  9.042  & 0.004 &  1.630 & 0.008 &  0.908 &  0.013 \\
6379.436 &  8.920  & 0.010 &  1.635 & 0.014 &  0.880 &  0.014 \\
6391.557 &  8.958  & 0.006 &  1.621 & 0.008 &  0.900 &  0.009 \\
6398.401 &  9.027  & 0.014 &  1.623 & 0.017 &  0.903 &  0.018 \\
6399.380 &  9.021  & 0.011 &  1.643 & 0.012 &  0.882 &  0.012 \\
6404.374 &  9.078  & 0.004 &  1.639 & 0.006 &  0.898 &  0.008 \\
6418.403 &  9.132  & 0.006 &  1.641 & 0.007 &  0.898 &  0.017 \\
6694.613 &  9.150  & 0.008 &  1.808 & 0.014 &  0.933 &  0.010 \\
6713.615 &  8.862  & 0.011 &  1.694 & 0.021 &  0.897  & 0.012 \\
6727.471 &  8.944  & 0.011 &  1.654 & 0.021 &  0.880  & 0.012 \\
6737.508 &  9.019  & 0.004 &  1.660 & 0.007 &  0.883  & 0.010 \\
6745.475 &  9.088  & 0.002 &  1.676 & 0.009 &  0.884  & 0.005 \\
6758.373 &  9.114  & 0.008 &  1.676 & 0.012 &  0.880  & 0.013 \\
6764.420 &  9.160  & 0.015 &  1.689 & 0.017 &  0.887  & 0.015 \\
6771.433 &  9.304  & 0.004 &  1.699 & 0.015 &  0.909  & 0.007 \\
6776.390 &  9.414  & 0.002 &  1.717 & 0.005 &  0.920  & 0.009 \\
6782.383 &  9.573  & 0.014 &  1.823  & 0.024 &  0.950  & 0.020 \\
6792.497 &  9.511  & 0.027  &     ...    &   ...      & 0.898  & 0.028 \\
6799.382 &  9.413  & 0.003  & 1.811 & 0.022  &  0.912 &  0.008 \\
6800.378 &  9.390  & 0.018  &  ...      &   ...     &  0.924  & 0.022 \\
6804.441 &  9.213  & 0.009  &  ...      &   ...     &  0.863  & 0.051 \\
6814.437 &  8.961  & 0.002  & 1.687  & 0.013 &  0.889  & 0.021 \\
\tableline
\end{tabular}
\end{center}
\end{table*}

\subsection{High-resolution spectroscopy}

 \subsubsection {Description of spectrum and line selection.}

\begin{figure*}  
 \resizebox{\hsize}{!}{\includegraphics{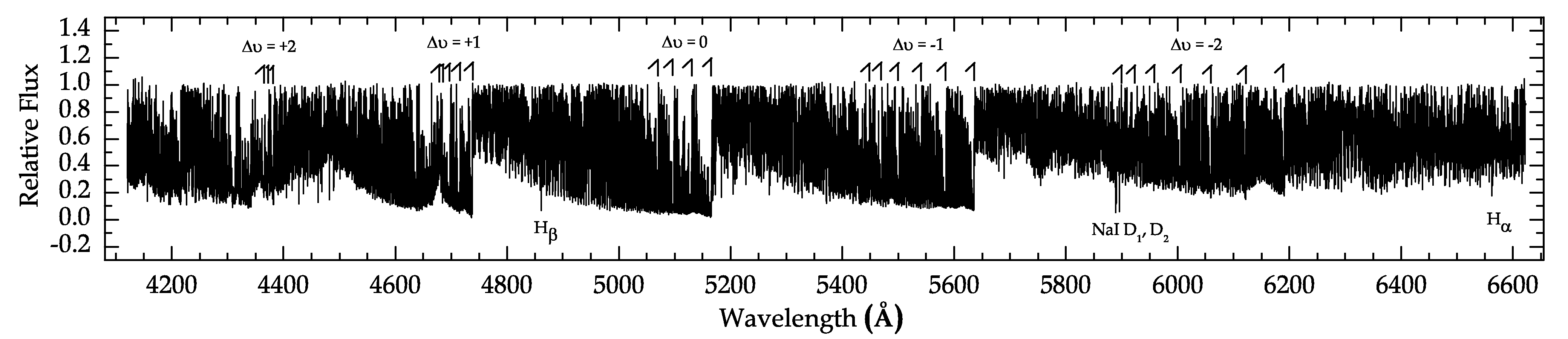}}
                              \caption{Normalized high resolution spectrum of HD\,112869 in the wavelength region from 4100 to 6650 \AA. The dominating $C_2$ Swan system $\Delta \upsilon = +2, +1, 0, -1, -2$ bandheads and a few prominent atomic lines are marked.}
  \label{fig6}
\end{figure*}

Strong absorption lines of carbon bearing molecules are dominating the spectrum of HD\,112869. The most prominent molecular features in the optical wavelength region are the $C_2$ Swan system band  heads (0,0) at 5165 \AA, (1,0) at 4740 \AA, and (0,1) at 5635 \AA\ (see Figure~\ref{fig6}). In addition, over the entire analyzed  wavelength region, strong lines of CN and CH molecules are evident, blending significantly the atomic absorption lines. However, in some wavelength regions, contamination from molecules is lower, e.g. redward of the Swan system bandheads.  
A qualitative  inspection of the observed high-resolution spectrum for HD\,112869 shows that, relative to the comparison stars, lines of the iron-peak and neutron-capture elements are much weaker and molecular lines with  isotopic $^{13}C$ are not clearly visible (e.g. Figures~\ref{fig1} and \ref{fig2}). The selection of relatively unblended atomic lines for abundance analysis was done using the synthesized molecular spectra over all the wavelength region employed.

\subsubsection{Atmospheric models}

\begin{figure}
\resizebox{\hsize}{!}{\includegraphics{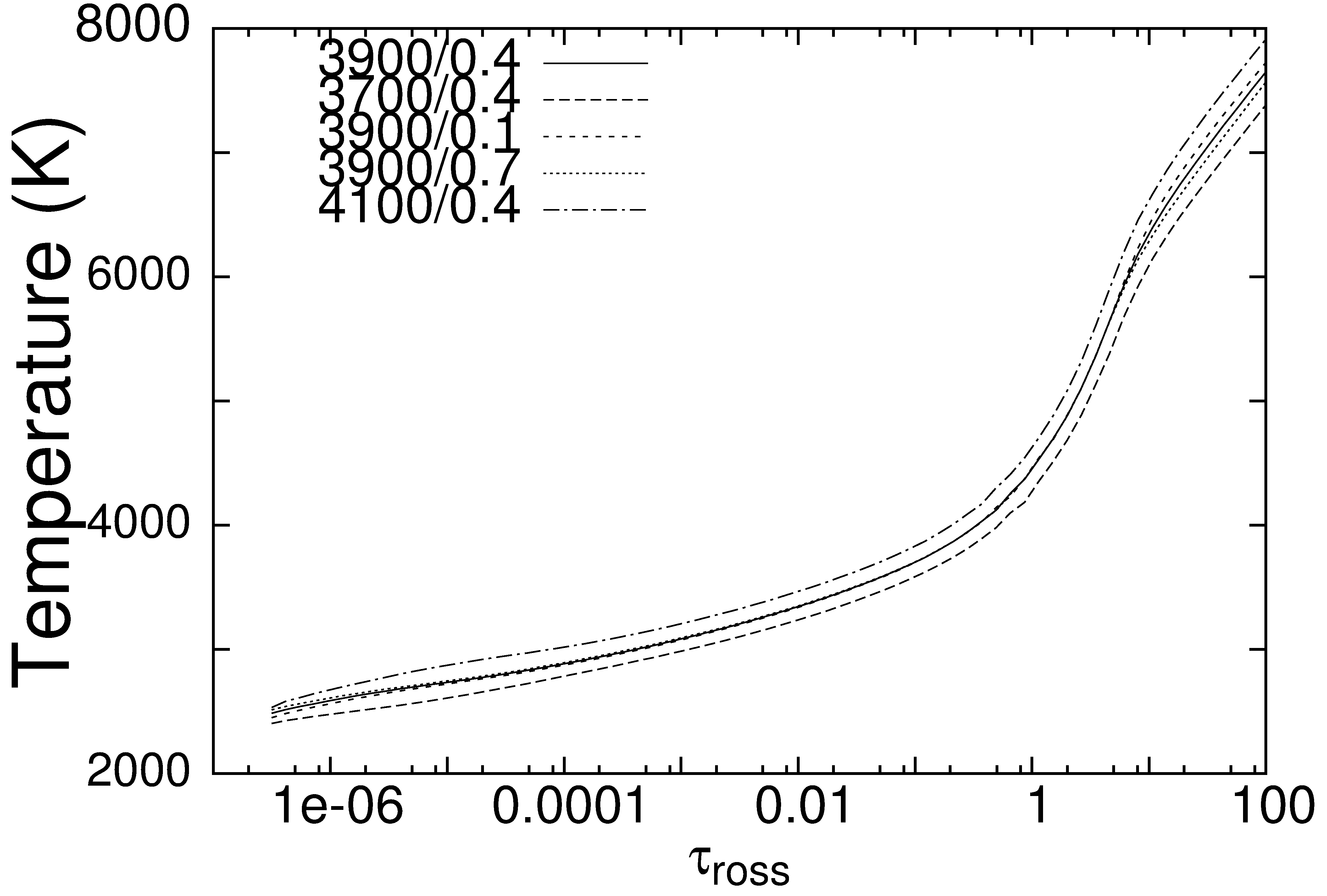}}
    \caption{Temperature structure of the final atmospheric model calculated for HD\,112869 (solid line), along with that for the modified models  ($\Delta T_{\rm eff}$ = $\pm$200~K, $\Delta \log g$ = $\pm$0.3 (cgs)) used for estimation of systematic errors.} 
     \label{fig7}
\end{figure}

The first iteration of abundance analysis was carried out using the atmospheric model calculated by T.Kipper with the code SCMARCS 21.1
for the published atmospheric parameters and abundances: $T_{eff}$ = 3700\,K, $\log g$ = 0.4 (cgs), $\xi_t$ = 4.0~km s$^{-1}$, [Fe/H] = $-$3.0, and 
$C/O$ = 1.07 \citep{kipper}. The next iterations were made using two atmospheric models calculated by U.G.J\o rgensen for $T_{eff}$ = 3800 K and a high carbon to oxygen ratio, $C/O$ =  2.5 and 26.3. The final atmospheric model  was calculated using the code SAM12, a modification of Kurucz's code ATLAS12  (Kurucz 2005), produced to calculate the atmospheric models of red giants of a given chemical composition \citep{pavlenko03}. The bound$-$free absorption caused by the C\,{\sc i}, N\,{\sc i}, and O\,{\sc i} atoms was added \citep{pz} to the continuum absorption sources included in ATLAS12. We also take into account the CIA (Collision Induced Absorption) -- absorption of the molecular complexes He$-H_2$ and $H_2-H_2$ induced by collisions which becomes an important source of opacity in the atmospheres of cool metal-poor stars \citep{borysow}. Molecular and atomic absorption in SAM12 is taken into account by using the opacity sampling technique \citep{sneen}. A compiled list of spectral lines includes the atomic lines from the VALD database \citep{VALD1, kupka} and the molecular lines CN, C$_2$, CO, SiH, MgH, NH, OH from the Kurucz database (Kurucz 1993). In addition, the absorption by the HCN bands was taken into account according to  \citet{harris, harris2}. We took into account also the absorption of the isomers HCN and NHC. The atmospheric model was calculated for the accepted atmospheric parameters, [Fe/H], and $\log (C/O)$. The model abundances were corrected according to the actual  chemical composition calculated for the HD\,112869 atmosphere during the subsequent iterations.  We adopted a high $^{12}C/^{13}C$ = 500 ratio in the calculations of the final atmospheric model. A temperature structure of the atmospheric models for carbon giants is less sensitive to $^{12}C/^{13}C$ \citep{pavlenko03}. The temperature structure of the final set of atmospheric models 
is shown in Figure~\ref{fig7}.

\subsubsection{Atmospheric parameters}

We adopted an effective temperature of 3900 K for the final atmospheric model, which is a mean value estimated using IRFM by three authors (see Section 1). The use of ionization balance for the calculation of surface gravity seems to be problematic because of significant effects of nonlocal thermodynamic equilibrium (NLTE)  for iron (see Section 3.3.5). We find $\log g$ lower than zero, if we estimate the gravity by assuming LTE and requiring the iron abundances derived by analysis of Fe\,{\sc i} and Fe\,{\sc ii} lines to be equal,  which contradicts the value of g calculated for $M_{bol} = -$3.35 mag assuming for HD\,112869 a mass of 1$M_{\sun}$, $\log$ g = 0.4 (cgs).  The microturbulent velocity was determined by forcing the abundances of individual lines to be independent of the equivalent width. The final atmospheric parameters used for abundance analysis are ($T_{eff}$, $\log g$, $\xi_t$) = (3900 K, 0.4 (cgs), 4.0 km s$^{-1}$).  Notice that the adopted values of effective temperature and surface gravity are in line with the overall relationship established for CEMP stars (see Figure 1 in \citet{bisterzo}.

\subsubsection{Spectral synthesis}

The abundances of carbon, nitrogen, and the ratio of carbon isotopes are estimated by synthesis of $C_2$  and CN spectra around the bandheads of Swan, Phillips,  and CN Red systems in a broad wavelength region from about 4700 to 9000 \AA,  using a standard technique adopted for the analysis of cool carbon-rich stars (see, for instance, \citet{zamora}). We concluded during the subsequent iterations  that the calculated $C_2$  and CN spectra reproduced the observed molecular features quite well near the bandheads  despite of the star's complicated spectrum. The abundances for other than carbon and nitrogen elements are calculated by synthesis of the profiles of less blended atomic lines (see Table~\ref{lines}), apart from strong bandheads where the number and intensity of molecular lines is lower. Unfortunately, the fit between the calculated and observed molecular spectrum for the final C and N abundances becomes worse apart from the molecular bandheads because of increased uncertainties of molecular data, especially for the $C_2$ Swan system. Therefore,  calculation of the abundances using atomic lines is a challenge for cool carbon-rich stars. Weak  atomic lines are detectable far from molecular bandheads where a correct estimation of molecular contribution is plagued due to lower quality of molecular data. In the case of poor reproduction of the observed molecular spectrum in the surroundings of selected atomic lines for the final C and N abundances, only synthesized atomic spectra were fitted to the observed spectrum and, in this way, an upper limit of abundance was derived (see Table~\ref{abundances}).

The synthetic spectra are calculated withs the code WITA \citep{ pavlenko97} and are convolved with the instrumental profile, for which we used a Gaussian with FWHM  from $\sim$0.20 to 0.30 \AA\ depending on the wavelength region. The system of continuum
opacity sources was used the same as for the model atmospheres calculations. As a primary source of atomic data the VALD database was used. The list of molecular lines was adopted from the Kurucz database \citep{kuruczs} and SCAN tape \citep{jorgensen}. The mean absolute and relative abundances, calculated using the final model and the adopted atmospheric parameters,  are given in Table 6, together with the internal random error ($\sigma$), calculated by a root mean squere method, and the number of lines (n) used in the analysis. The standard notations are adopted everywhere.\footnote{[A/B] = $\log\,(N_A/N_B)_{\star}$ - $\log\,(N_A/N_B)_{\sun}$, where $N_A$, $N_B$  is the number density of an element A and B. \\ $\log\,\epsilon(A)$ = $\log\,(N_A/N_H)$ + 12.00, where $N_H$ is the number density of hydrogen.}  
The resulting relative abundances normalized by the solar-system abundances are calculated using the chemical composition for the solar photosphere provided by \citet{asplund2}.
 Figures~\ref{fig8}$-$\ref{fig21} illustrate synthesis of representative wavelength regions and atomic lines.

The systematic errors in the abundances, produced by uncertainties in $T_{eff}$ ($\pm$200~K), $\log g$ ($\pm$0.3 dex), and $\xi_t$ ($\pm$0.5~km s$^{-1})$, would lead to errors less than 0.3 dex (except for nitrogen). The uncertainty in the continuum normalization was estimated to be within 2\% over the most of the wavelenght regions employed. 
The atmosphere of HD\,112869 is warmer and the spectrum is less crowded in comparison with those observed for N type carbon stars. In most of the modelled regions the S/N ratio is high and the continuum was clearly identified in the spectrum. The synthesized atomic and molecular spectra helped us to define short wavelength regions in the spectrum which are free of absoption lines. The spectrum was normalized using a low-order polynomial fit to the selected regions. In the case of doubt, some of the regions were omitted and the rest ones were fitted again with the goal to examine the uncertainty of normalization. On the other hand, the carbon abundance was calculated using $C_2$  lines in the broad wavelength region from about 4700 to 9000 \AA\ (see Section 3.3.6).  We have not found any trend of the calculated carbon abundance from the wavelength of employed region, $\ log\,\epsilon(C)$ = 8.3 $\pm$ 0.1 dex. Thus, the systematic error in the spectrum normalization should be relatively low.  
Blueward of $C_2$ bandheads, the uncertainty exceeds 5\%, a value typical for cool carbon stars (see, for instance, \citet{abia02}. The systematic errors in the calculated abundances ($\Delta \log \epsilon$(X)) are estimated on the basis of abundances derived using the modified atmospheric models (see Figure~\ref{fig7}).  The conclusion is that the error due to uncertainty in the adopted temperature is  the largest for the nitrogen abundance and decreases for the abundances of iron peak  and s-process elements:  $\Delta \log \epsilon$(C) $\sim$0.2 dex, $\Delta \log \epsilon$(N) $\sim$ 0.5 dex, $\Delta \log \epsilon$(Mg) $\sim$0.15 dex, $\Delta \log \epsilon$(Ti) $\sim$0.25 dex,  $\Delta \log \epsilon$(Fe) $\sim$0.15 dex, and $\Delta \log \epsilon$(s-process) $<$0.1 dex. The systematic errors due to uncertainty in the adopted gravity are larger for the abundances derived by analysis of the ionized lines. We conclude, that these errors are less than 0.1 dex. The systematic errors arising from uncertainty in the adopted microturbulent velocity are larger for the elements represented by relatively strong lines. For example, the magnesium abundance was calculated on the basis of lines with the equivalent widths between 300 and 400 m\AA. However, these systematic errors are less than $\sim$0.1 dex for the lines used. The systematic errors due to uncertainties in the continuum normalization are  larger for the abundances derived by analysis of weak lines and these errors can reach 0.1 dex. The systematic error in the derived $^{12}C/^{13}C$ ratio, produced by an uncertainty of 2\% in the continuum normalization over the wavelength region around $C_2$ (1,0) bandhead, was found to be $\pm$300. The root-sum-squared (RSS) uncertainties calculated taking into account all sources of errors are less than $\sim$0.4 dex, what is a typical total uncertainty in abundance analysis of cool carbon stars \citep{zamora}.

\begin{table}
\begin{center}
\caption[]{List of the lines for neutron-capture elements, synthesized in the spectrum of HD\,112869. Wavelengths, excitation potentials, and gf-values are given.\label{lines}}
\begin{tabular}{cccr}
\tableline\tableline
 Wavelength & Species  &     LEP    &   $\log$(gf)   \\
      ( \AA\ )      &              &    (eV)     &       \\
\tableline

 4077.709 &  Sr\,{\sc ii} & 0.00  &  0.167    \\
 4086.709 &    La\,{\sc ii} & 0.00 & -0.070  \\
 4220.660  & Sm\,{\sc ii} & 0.544 & -0.440    \\
 4900.120  & Y\,{\sc ii}     &   1.03 &  -0.09    \\
 4934.076 &   Ba\,{\sc ii}  &   0.00 &  -0.15       \\
 5200.406 &   Y\,{\sc ii}  & 0.992  &  -0.57   \\
  5205.724 &  Y\,{\sc ii}  & 1.033 &  -0.340   \\
 5249.576 &   Nd\,{\sc ii}   &  0.976  &   0.200      \\
 5255.502&    Nd\,{\sc ii}   & 0.205   & -0.670    \\
 6141.713 &  Ba\,{\sc ii}  &  0.704 &  -0.076     \\
 6645.094  &   Eu\,{\sc ii}  & 1.380  & -0.162    \\
 6645.114  &   Eu\,{\sc ii}   & 1.380  & -0.200    \\
\tableline
\end{tabular}
\end{center}
\end{table}

\begin{table}
\begin{center}
\caption[]{Averaged absolute and relative abundances for HD\,112869, along with the random error and the number 
of lines used in the synthesis.\label{abundances}}
\begin{tabular}{crcccr}
\tableline\tableline
X     &    $\ log\,\epsilon(X)$   &   n   &   $\sigma$    &    [X/H]   &   [X/Fe\,II]    \\
\tableline
C            &  8.30     & $C_2$ &  0.1   & -0.13   &  +2.17      \\
N            &  6.55     &  CN      &  0.2   & -1.28   &  +1.02     \\
Mg\,{\sc i}      &  4.6    &  3   &  0.3   & -3.00   &  -0.7   \\ 
Ti\,{\sc i}     &  2.7    &     3   &   0.3       & -2.23   &    +0.1    \\
Ti\,{\sc ii}     &  3.1    &  2      &          & -1.85   &    +0.5    \\
Cr\,I    &  2.4    &      3     &    0.3     & -3.29   &  -1.0     \\
Fe\,{\sc i}        &  4.9    &    14    &   0.2 & -2.60   & -0.3        \\ 
Fe\,{\sc ii}        &  5.2   &   6       &  0.2   & -2.30   &     ...            \\
Sr\,{\sc ii} &  $\la$ -1.1   &      1     &     & $\la$ -3.97 & $\la$-1.7      \\
Y\,{\sc ii}  &  $\la$ -0.1   &  3       & 0.2       &  $\la$ -2.27  & $\la$ 0.0    \\
Ba\,{\sc ii} & $\la$ -0.4  & 2        &     & $\la$ -2.6   &  $\la$-0.3    \\
La\,{\sc ii} & $\la$ -0.4 &    1     &               &   $\la$-1.55  &   $\la$+0.8     \\
Nd\,{\sc ii}    & $\la$ -0.2  &   2   &     & $\la$ -1.67   & $\la$+0.7      \\
Sm\,{\sc ii}   &  $\la$-0.8          &    1    &          &     $\la$ -1.80 & $\la$+0.5  \\
Eu\,{\sc ii}    &   $\la$ -1.0  &    1     &      &  $\la$-1.52    &  $\la$+0.8   \\
\tableline
\end{tabular}
\end{center}
\end{table}

\subsubsection{Iron}

\begin{figure*}  
 \resizebox{\hsize}{!}{\includegraphics{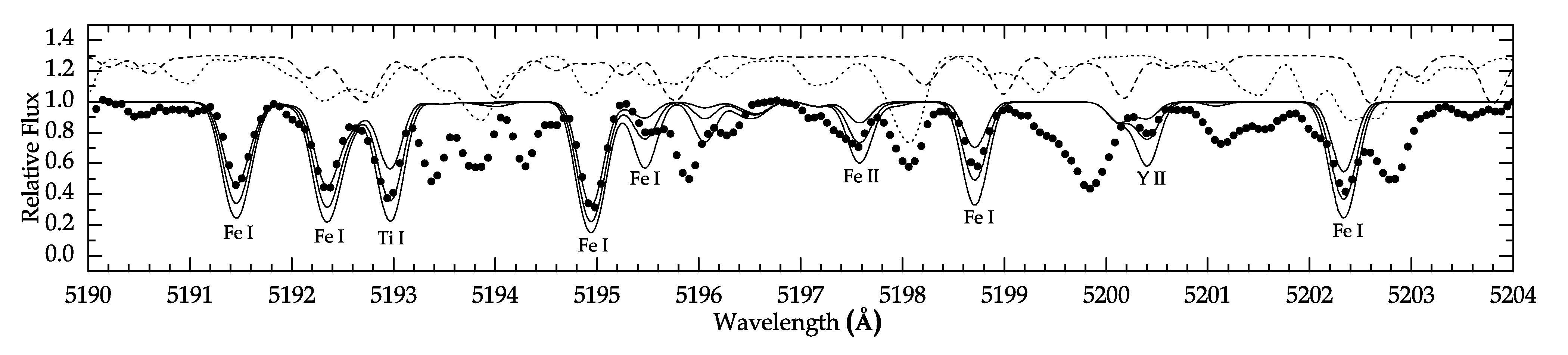}}
                              \caption{Observed spectrum of HD\,112869 (filled circles), along with the synthesized atomic (solid line) and molecular (CN: dotted line; $C_2$: dashed line) spectra, in the wavelength region less blended by molecules, illustrating the estimation of metallicity. The synthesized profiles of atomic lines are shown for three abundances: $\ log\,\epsilon(Fe)$ = 5.2 $\pm$ 0.5 dex, $\ log\,\epsilon(Ti)$ = 2.7 $\pm$ 0.5 dex, and $\ log\,\epsilon(Y)$ = --0.1 $\pm$ 0.5 dex. Synthesized molecular spectra are calculated for the final carbon and nitrogen abundances.}
  \label{fig8}
\end{figure*}

\begin{figure*}  
 \resizebox{\hsize}{!}{\includegraphics{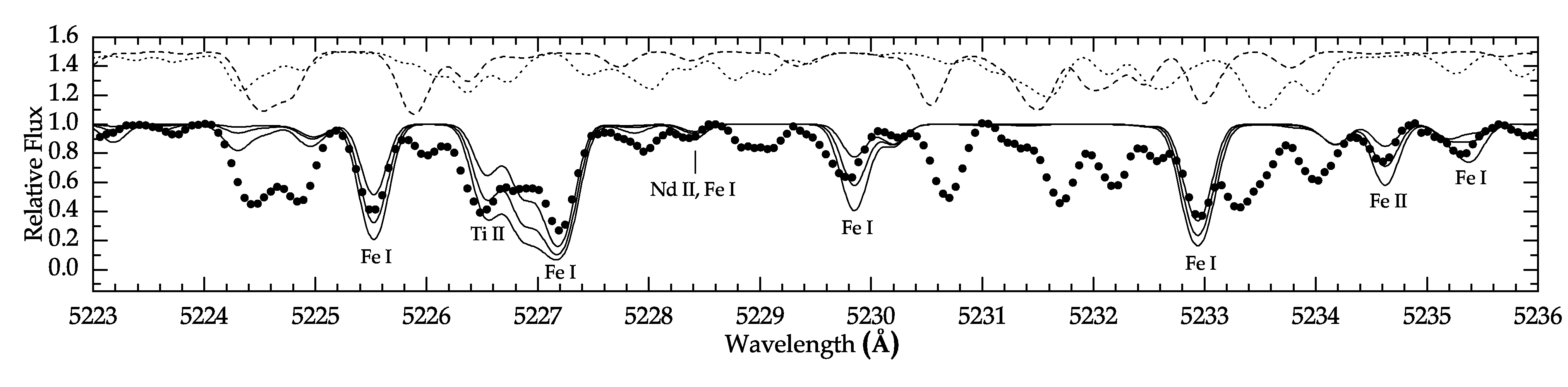}}
                              \caption{Same as Fig.~\ref{fig8}, but  in other spectral region.}
  \label{fig9}
\end{figure*}

A limited number of relatively clean iron lines was selected for HD\,112869. Using six Fe\,{\sc ii}  lines, we calculated the star's mean metallicity, [Fe/H] = $-$2.3, however, the neutral lines of Fe\,{\sc i}  provide lower metallicity, [Fe/H] = $-$2.6 dex, on average (see Figures~\ref{fig8} and ~\ref{fig9}).  It seems likely that neutral atomic lines suffer from NLTE  effects; these are significant at low metallicity and high luminosity. \citet{shchukina} have shown that for metal-poor stars the NLTE abundance corrections for Fe\,{\sc i} can reach about 0.9 dex. As argued by many authors, Fe\,{\sc ii} lines are more reliable abundance indicators than Fe\,{\sc i} lines, since such lines are nearly independent on the departures from NLTE and the temperature structure of model atmospheres. Thus, we conclude that the metallicity of HD\,112869 is higher than estimated before, [Fe/H] = $-2.3 \pm$0.2 dex, assuming that the lines of ions are the safest abundance indicators. However, the iron group element chromium seems to be more depleted than iron, [Cr/Fe] $\simeq -$1.0 dex, as based on neutral chromium lines. Note that the iron peak elements chromium and manganese are underabundant in most of the analyzed CEMP stars and the [Cr/Fe] ratio decreases at lowest Fe abundances \citep{norris01, allen}.

\subsubsection{Carbon} 

\begin{figure*}  
\resizebox{\hsize}{!}{\includegraphics{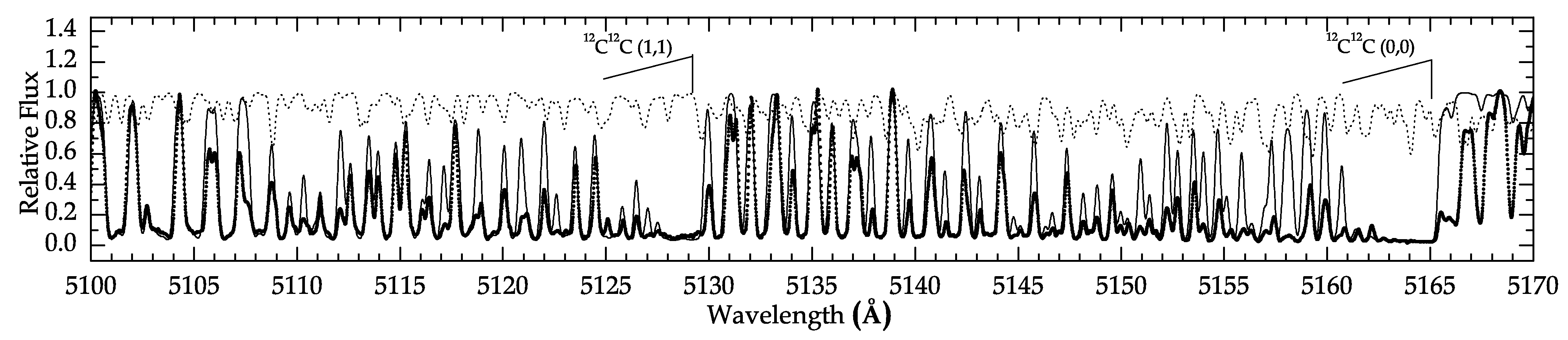}}
                              \caption{The observed spectrum of HD\,112869 (solid dots), along with synthesized
$C_2$ (solid line) and CN (dotted line) spectra, in the wavelength region dominated by Swan system $\Delta \upsilon$ = 0 lines. The calculated spectra are shown for the final abundances, $\ log\,\epsilon(C)$ = 8.3 and $\ log\,\epsilon(N)$ = 6.55.}
  \label{fig10}
\end{figure*}

\begin{figure*}  
\resizebox{\hsize}{!}{\includegraphics{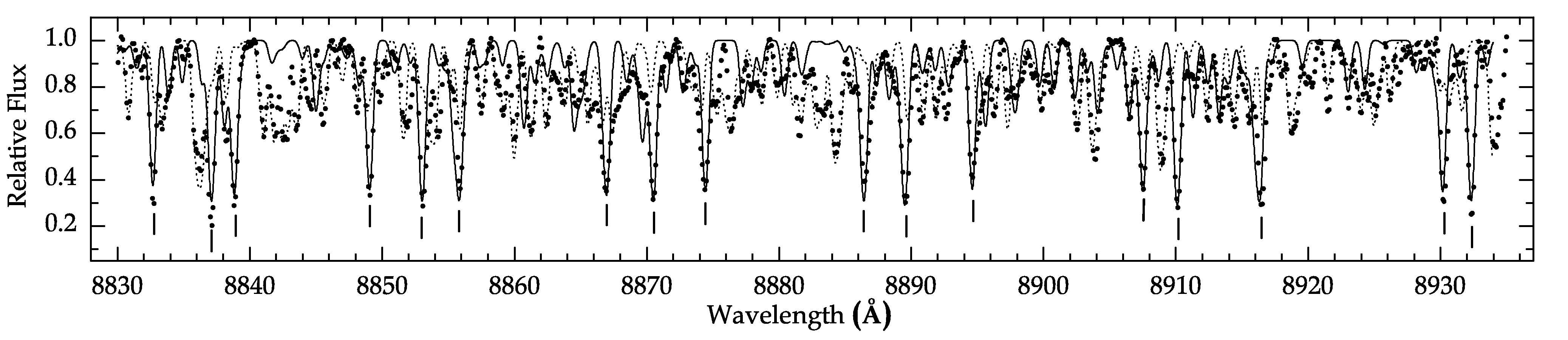}}
                              \caption{Same as Fig.~\ref{fig10}, but in the wavelength region dominated by $C_2$ Phillips system lines. The synthesized $C_2$ and CN spectra are shown for the final abundances.}
  \label{fig11}
\end{figure*}

\begin{figure}  
\resizebox{\hsize}{!}{\includegraphics{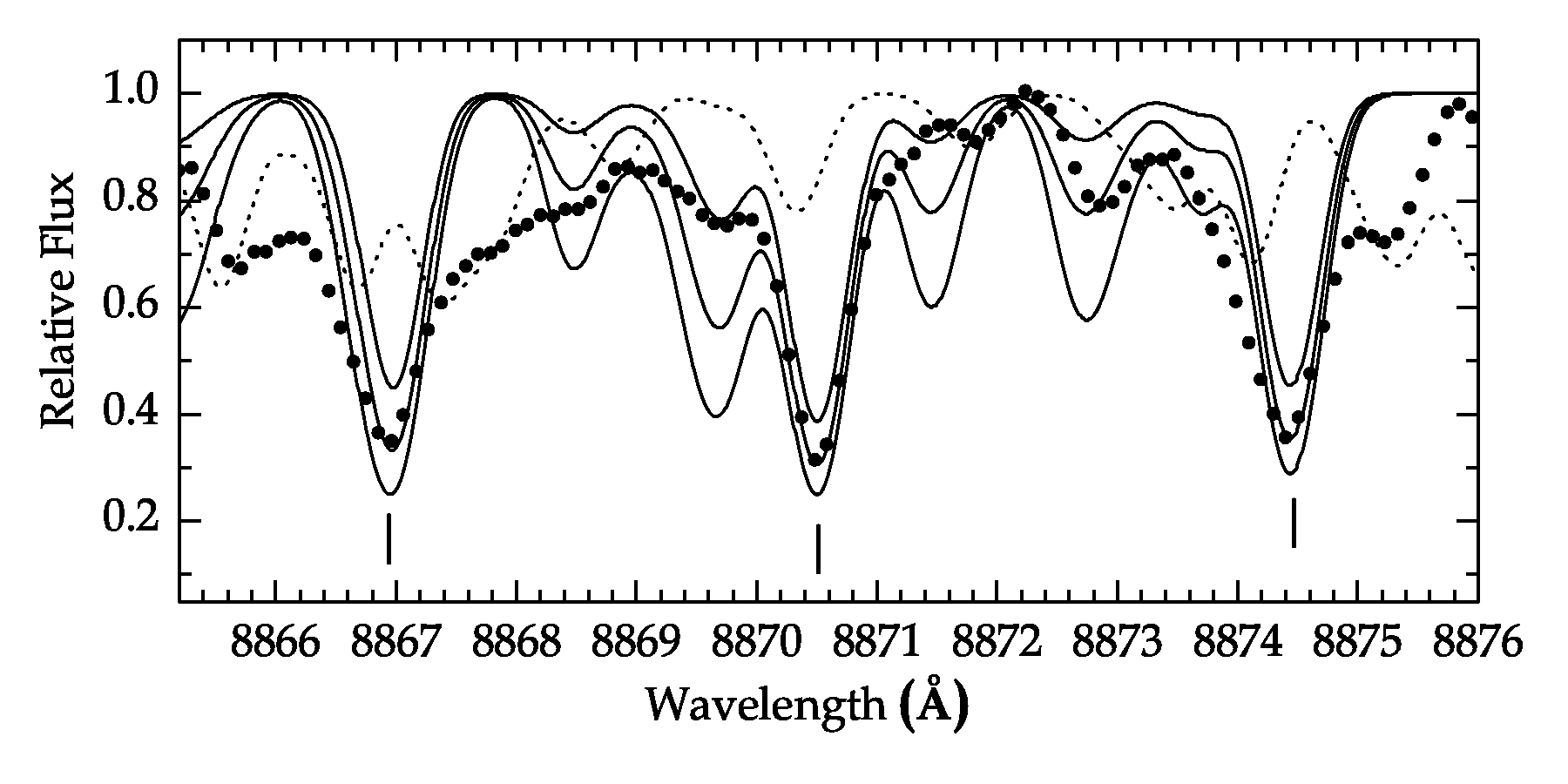}}                               
\caption{The observed spectrum of HD\,112869 (filled circles), along with synthesized $C_2$ (solid line) and CN (dotted line) spectra, in the wavelength region dominated by 
Phillips system lines, illustrating the uncertainty in the derived carbon abundance. The synthesized  $C_2$ spectra are given for three carbon abundances, $\ log\,\epsilon(C)$ = 8.3 $\pm$ 0.2 dex. The synthesized CN spectra are shown for the final C and N abundances.}
\label{fig12}
\end{figure}

Carbon abundance was derived from the $C_2$ Swan and Phillips system lines synthesized in the wavelenght region from about 4736 to 6200 \AA\ and in the near infrared.  Figures~\ref{fig10} and ~\ref{fig11} illustrate the fit between the observed and calculated spectrum  in two wavelengths regions  - around $C_2$ Swan system  bandheads (0,0) and (1,1) shortward of 5170 \AA\ and around Phillips system lines in the near infrared  from 8830 to 8931 \AA. The given synthetic spectra are generated using the final atmospheric parameters adopted for the high carbon abundance, $\ log\,\epsilon(C)$ = 8.3. The $C_2$  Swan system bandheads are saturated and the sensitivity of the calculated spectra to the adopted carbon abundance near the bandheads is low. However, the sensitivity increases for weaker lines between the bandheads.
In Figure~\ref{fig12} we illustrate the sensitivity of the calculated spectra from the adopted carbon abundance. The mean carbon abundance was found to be high in the atmosphere of HD\,112869, $\ log\,\epsilon(C)$ = 8.3 $\pm$ 0.1 dex.

\subsubsection{Nitrogen} 

\begin{figure*}
\includegraphics[width=14cm]{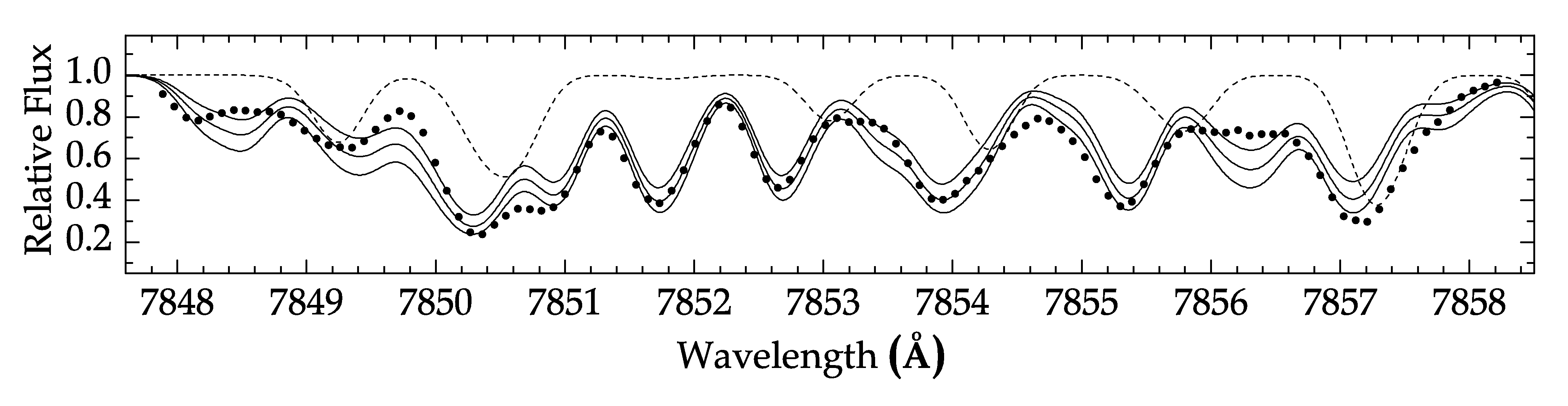}   
\caption{The observed spectrum of HD\,112869 (filled circles), along with synthesized
CN (solid line) and $C_2$ (dashed line) spectra, in the wavelength region dominated by CN Red system lines, illustrating the uncertainty in the derived nitrogen abundance. The synthesized CN spectra are given for three nitrogen abundances,  $\ log\,\epsilon(N)$ = 6.55 $\pm$ 0.2 dex. The adopted carbon abundance is $\ log\,\epsilon(C)$ = 8.3.}
     \label{fig13}
\end{figure*}

The nitrogen abundance was calculated by spectral synthesis of CN Red system lines in the wavelength region from about 7800 to 8100 \AA. The sensivity of the synthesized CN spectrum to the adopted abundance of nitrogen is  lower than in the case of  $C_2$ lines to the carbon abundance. Therefore, the uncertainty in the fitting of the synthetic CN spectrum to the observed one is larger. The best fit between the observed and synhesized spectra was found for the nitrogen abundance  $\ log\,\epsilon(N)$ = 6.55. An extract of the spectra illustrating the synthesis of CN Red system  lines is shown in Figure~\ref{fig13}.

\subsubsection{$^{12}C/^{13}C$ ratio} 

\begin{figure*}   
\resizebox{\hsize}{!}{\includegraphics{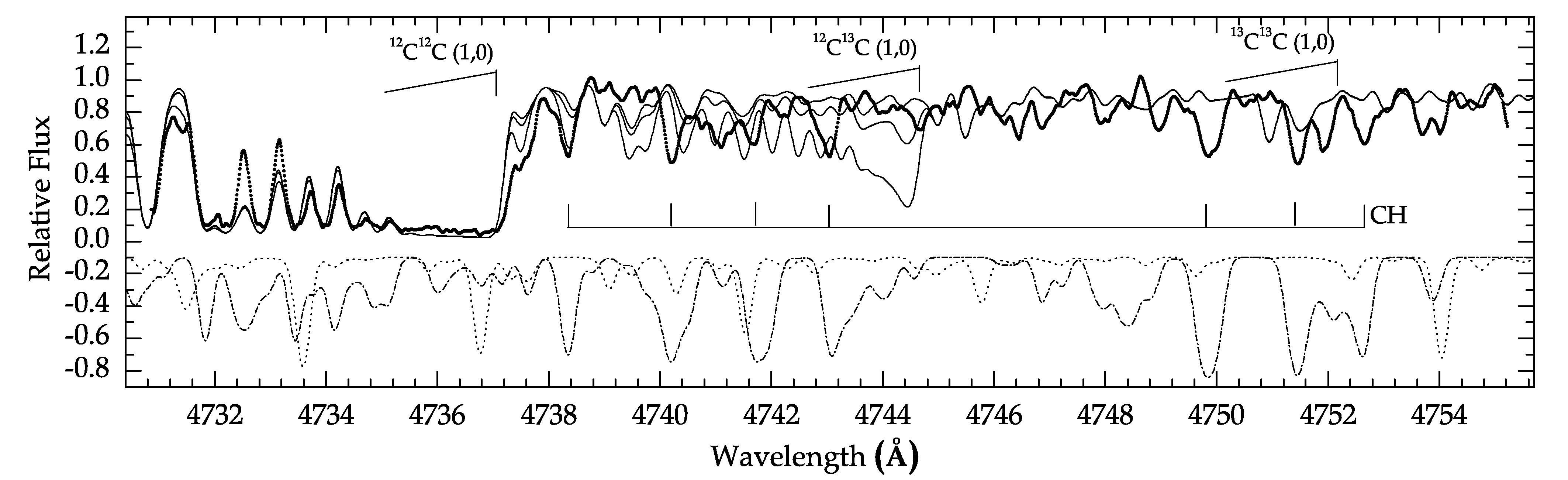}}
                         \caption{The observed spectrum of HD\,112869 (solid dots), along with synthesized
spectra, in the wavelength region of $C_2$ Swan system (1,0) bandheads. Synthesized $C_2$ spectra (solid lines) are shown for the  final carbon abundance and three isotopic ratios, $^{12}$C/$^{13}$C= 90, 500, 1500. Synthesized CH (dotted-dashed line) and atomic (dotted line) spectra are presented at the bottom. CH lines identified in the observed spectrum are marked by vertical ticks.}
  \label{fig14}
\end{figure*}

\begin{figure*}  
\resizebox{\hsize}{!}{\includegraphics{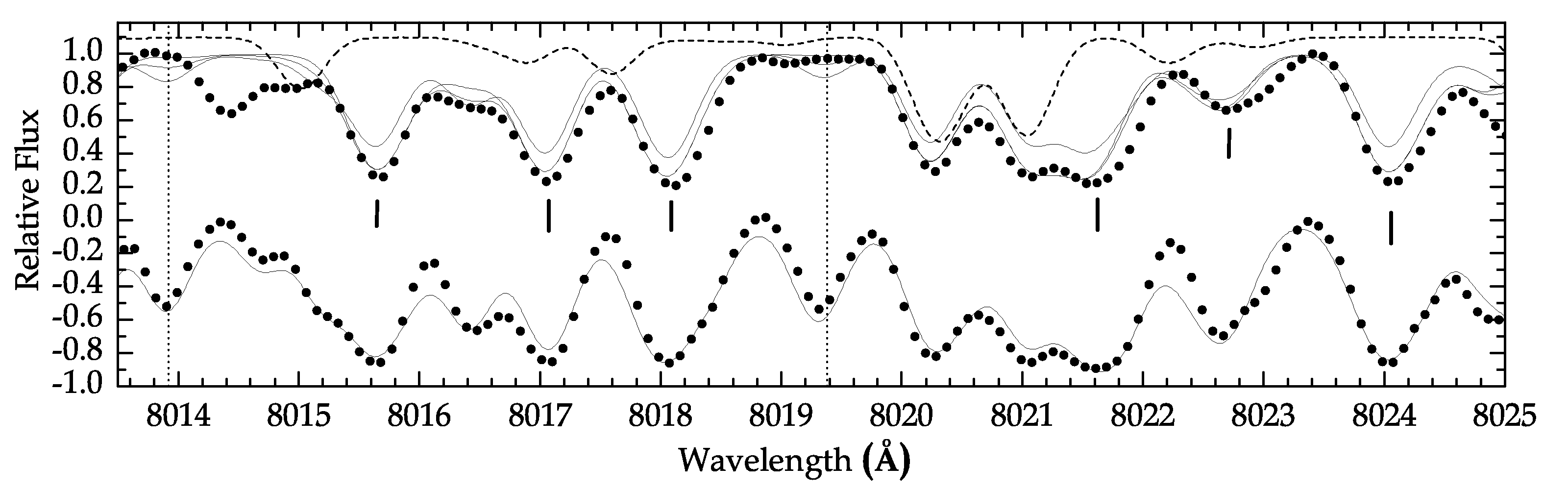}}                         
              \caption{The observed spectrum of HD\,112869 (filled circles on the top), along with synthesized
spectra, in the wavelength region dominated by CN Red system lines. Three synthesized CN  (solid line) and one $C_2$ (dashed line) spectra are shown for  the final nitrogen and carbon  abundances and the isotopic ratios $^{12}$C/$^{13}$C= 5, 90, 500. The positions of $^{12}CN$ and $^{13}CN$ lines are indicated by vertical ticks and dotted vertical lines, respectively. The observed (filled circles) and synthesized spectra for the comparison star HD\,92055,  calculated using the model parameters given in Table~\ref{basic_dat}, are presented at the bottom for comparison purposes.}
 \label{fig15}
\end{figure*}

An inspection of the spectrum of HD\,112869 in wide wavelength region gives evidence that isotopic lines are too weak to be clearly detected in the crowded spectrum, therefore, only the lower limit of   $^{12}C/^{13}C$ ratio can be determined. The $C_2$ (1,0) isotopic bandheads at 4744 and 4752 \AA\ synthesized for $\ log\,\epsilon(C)$ = 8.30 and three  isotopic ratios, $^{12}C/^{13}C$ = 90, 500, and 1500, are given in Figure~\ref{fig14}. Moderatelly strong features of unknown nature, observed by \citet{aoki} in the wavelength region between 4738 and 4754 \AA, were identified by CH lines, however, the synthesized CH spectrum is too strong for the final carbon abundance. Overestimated oscillator strengths are suspected. CH lines are blending significantly the $^{13}C^{13}C$  (1,0) bandhead, therefore, the $^{12}C/^{13}C$ limit was set using the $^{12}C^{13}C$  (1,0) bandhead. We conclude, that the isotopic ratio is extremely high in the atmosphere of HD\,112869, $^{12}C/^{13}C  \gtrsim$ 1500. In addition, the isotopic lines of CN Red system are examined in the wavelength region from 7800 to 8100 \AA. We selected less crowded wavelength regions with a certain continuum definition to estimate a reliable limit for the isotopic ratio. The results of spectral synthesis in one narrow wavelength region for three isotopic ratios are shown in Figure~\ref{fig15}. The isotopic $^{13}C$N lines are not detected certainly in the spectrum of HD\,112869,  in agreement with the extremely high over 500 isotopic ratio.

\subsubsection{Oxygen} 

Spectroscopic estimation of the oxygen abundance in the optical wavelength region remains a challenge for cool carbon-rich stars. The most reliable oxygen abundances come from the forbidden [OI]  line at 6300 \AA. However, this line should be quite weak at low temperature and metallicity. Most of the previous high-resolution calculations of oxygen for CEMP stars were based on measurements of the lines of infrared triplet at 7771.94, 7774.17, and 7775.39  \AA, which are strongly affected by NLTE effects \citep{asplund}.  We synthesized the spectral region around the infrared triplet using the final atmospheric model for three oxygen abundances, $\ log\,\epsilon(O)$ = 0.0, 6.7, and 8.7 dex. The conclusion is that at the position of the infrared triplet dominate CN lines. The oxygen lines  are too weak to be detected  in the spectrum of HD\,112869 even for the solar abundance. Therefore, we adopted for HD\,112869 the oxygen abundance $\ log\,\epsilon(O)$ = 7.2 dex according to the relationships [O/Fe] vs. [Fe/H] \citep{aoki04} and [O/Fe] vs. [C/Fe] \citep{kennedy} observed for CEMP stars : [O/Fe] $\simeq$ +0.8 for [Fe/H] = $-$2.3 dex.

\subsubsection{Magnesium}

\begin{figure*}
\resizebox{\hsize}{!}{\includegraphics{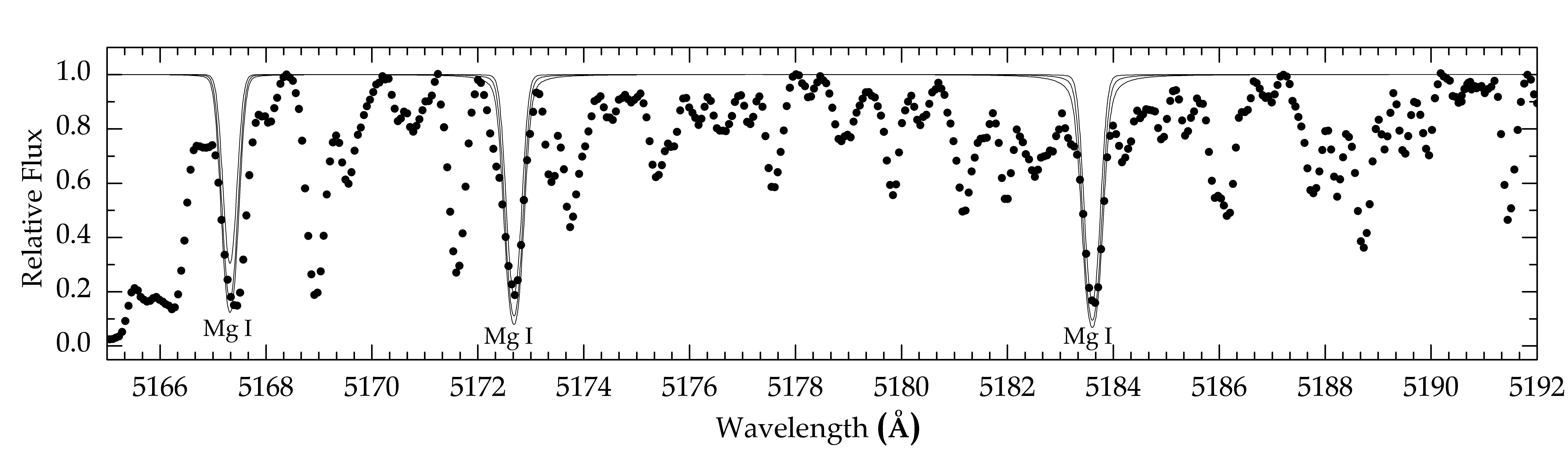}} 
\caption{The observed spectrum of HD\,112869 (filled circles), along with calculated
profiles for magnesium triplet. Mg\,{\sc i} lines  are synhesized for three magnesium abundances, $\ log\,\epsilon(Mg)$ = 4.6, 5.3, and 5.7 dex (solid lines).}
    \label{fig16}
\end{figure*}

Magnesium is a key element for examination of $\alpha$-process nucleosynthesis in the Galaxy and Mg/Fe ratio carries information about the initial mass function. The Mg\,{\sc i} b triplet is located redward of the $C_2$ (0,0) Swan system bandhead at 5165 \AA\ in the wavelenghts region less blended by molecular lines and with a certain continuum definition.   
Mg\,{\sc i} lines at 5167.322, 5172.685, and 5183.604 \AA\  are synthesized to estimate  the magnesium abundance. The line at 5167 \AA\ is a significant blend, while the rest two lines are almost clean. Spectral synthesis  yields the abundance  $\ log\,\epsilon(Mg) \simeq$ 4.6,  with a reasonable fit for the line at 5172 \AA\ with an equivalent width of about 360 m\AA\  and the line at 5183 \AA\ with an equivalent width of about 395 m\AA\ (see Figure~\ref{fig16}).  Note, that the list of lines adopted for spectral synthesis contains only Mg lines, and the weak
blends inside  the profiles for lines at 5171 \& 5183 \AA\ were not taken into account. Thus, the magnesium abundance seems to be lower in the atmosphere of HD\,112869 in comparison with the average value found for CEMP stars.  A moderate magnesium overabundance, [Mg/Fe] $\simeq$ + 0.4 dex, and the absence of a significant trend of $[\alpha/Fe]$ vs.  [Fe/H] were found for a large sample of CEMP stars (Allen et al. 2012 and references therein). The calculated NLTE corrections for Mg\,{\sc i} lines (including Mg\,{\sc i} b triplet) should not exceed +0.2 dex for the atmospheric parameters similar to those of HD\,112869 \citep{merle, mashonkina}. The abundance of another $\alpha$-element, titanium, was found to be moderately enhanced relative to iron, [Ti/Fe] = +0.4 dex, with a tendency to provide a lower abundance from neutral lines. 

Possible sources of uncertainties involved in abundance analysis are analyzed to clarify the reason of low Mg abundance found for HD\,112869. The lines of Mg triplet are moderately strong and the calculated abundance can be affected by the error due to uncertainty in the adopted microturbulent velocity. The profiles of Mg triplet were synthesized for reduced microturbulence, $\xi_t$ = 3.0~km s$^{-1}$. In this case, the Mg abundance is higher only by about 0.1 dex. The error in the abundance, produced by the uncertainty in the adopted temperature, is $\pm$0.16 dex. The wavelength region redward of the $C_2$ (0,0)  bandhead is free from strong molecular lines and the accuracy of continuum localization was estimated to be better than 2\%. We concluded during subsequent iterations that, the local continuum should be by about 10\% higher than the adopted one in order to have the magnesium abundance ($\log\,\epsilon(Mg)$ = 5.7) observed for typical CEMP stars, [Mg/Fe] = +0.4 dex. In that case, however,  the continuum should be scaled up for all lines in the wavelength region near Mg I triplet (e.g.  Figures~\ref{fig8} and \ref{fig9}), what is unlikely. To characterize a total error of the calculated magnesium abundance, $\pm$0.4 dex, the RSS  uncertainty was estimated taking into account the contribution from both the random and systematic errors.  Weak Mg lines should be employed to confirm our result.

\subsubsection{Neutron--capture elements}

\begin{figure*}
\includegraphics[width=14cm]{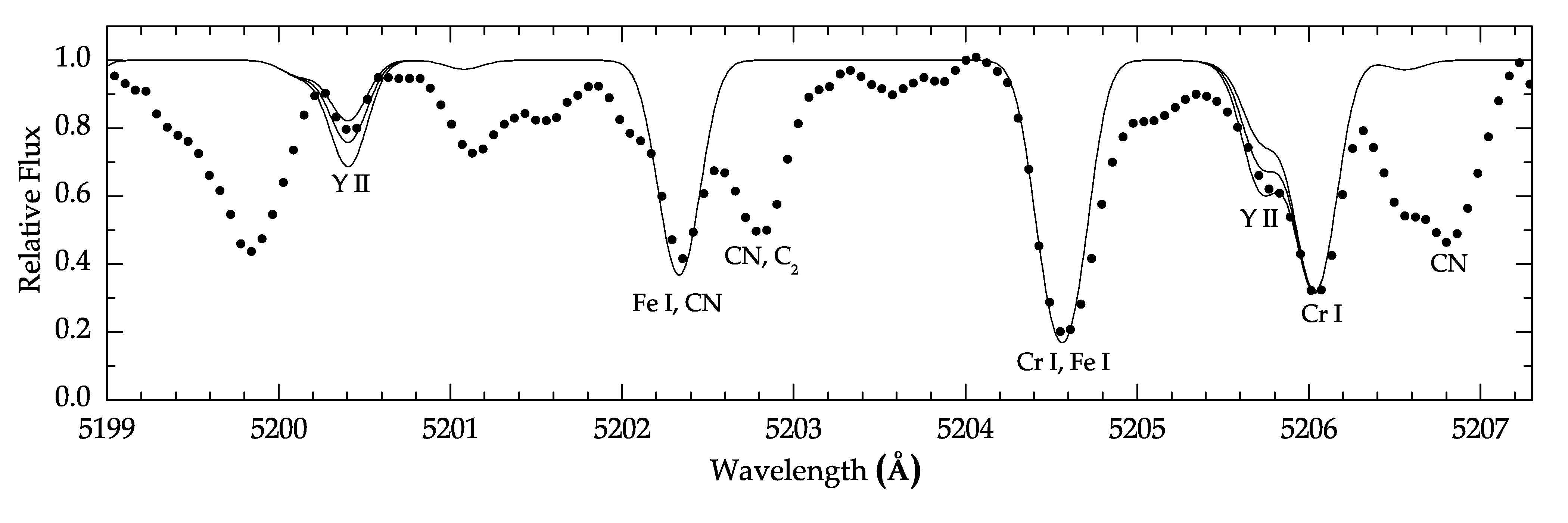}   
\caption{The spectrum of HD\,112869 (filled circles), along with synthesized atomic spectra around Y\,{\sc ii} lines at 5200.406 and 5205.724 \AA. The synthesized spectra are shown for three yttrium abundances, $\ log\,\epsilon(Y)$ = -0.1 $\pm$ 0.2 dex.}
    \label{fig17}
\end{figure*}

\begin{figure}
\resizebox{\hsize}{!}{\includegraphics[angle = 0]{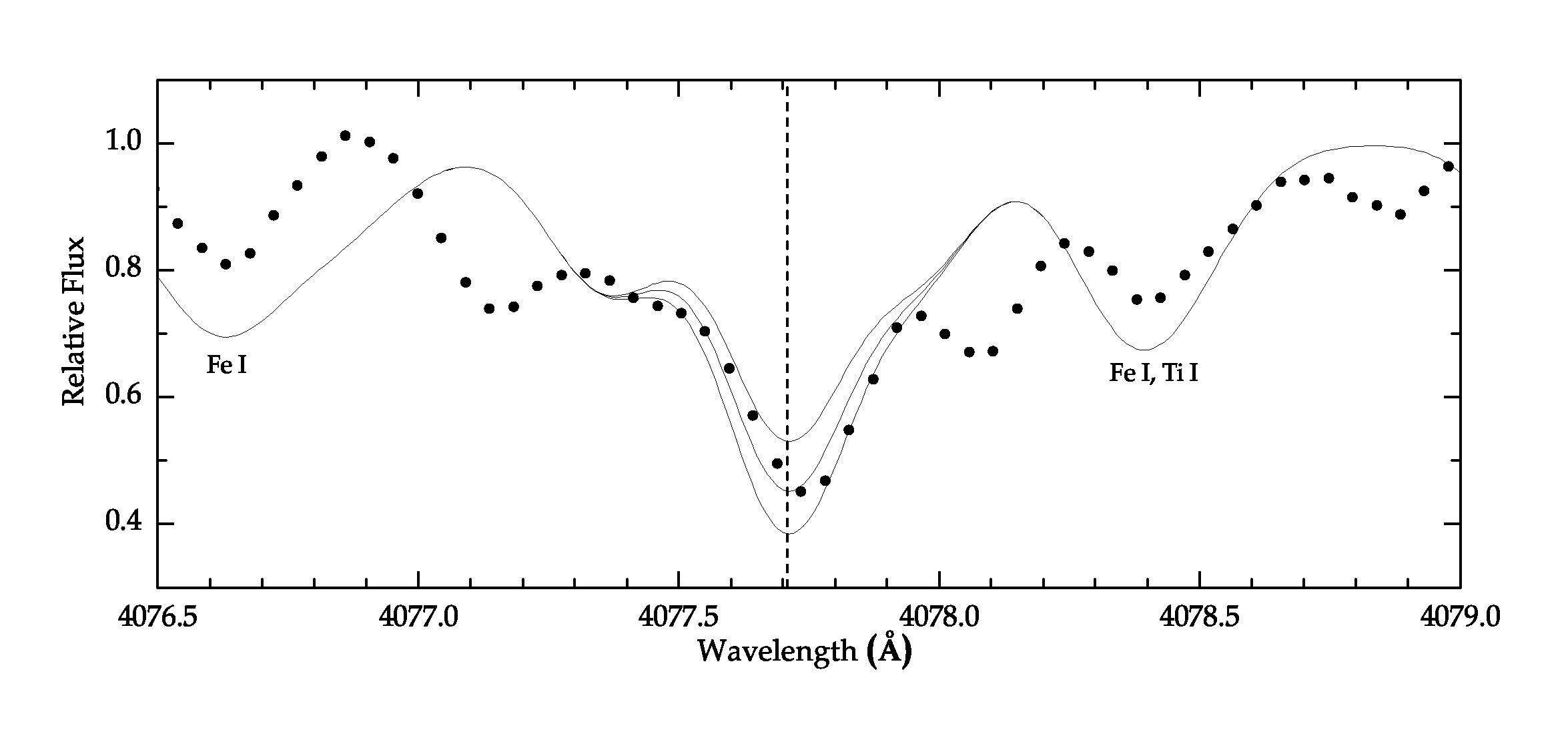}}
    \caption{The spectrum of HD\,112869 (filled circles), along with synthesized atomic spectra around Sr\,{\sc ii} line at 4077.709 \AA. The synthesized spectra are shown for three strontium abundances, $\ log\,\epsilon(Sr)$ = -1.1 $\pm$ 0.3 dex.} 
     \label{fig18}
\end{figure}

The elements formed mainly by the s-process main component can be divided into two groups: light s-process (ls)  elements around the magic neutron number 50 and heavy s-process (hs) elements around the magic neutron number 82. An enhancement of the light $s$-process elements in the atmosphere of HD\,112869 relative to the solar values was not confirmed. We selected two Y\,II lines for the abundance estimation in the wavelength region redward of the $C_2$ Swan system bandhead with a certain definition of the continuum level. The lines at 5200.406  and  5205.724 \AA\  give a similar yttrium abundance within uncertainties (see Figure~\ref{fig17}). The chromium abundance was iterated to be $\ log\,\epsilon(Cr)$ = 2.35 using the Cr\,I line at 5206.023 \AA\ which is blending with the yttrium line. The third Y\,II line at 4900.12 \AA\ confirms the low yttrium abundance and an average relative abundance normalized to the iron abundance was calculated, which appears to be near zero, [Y/Fe] $\simeq 0.0$ dex.  Unfortunatelly, we were not able to reproduce well $C_2$ and CN spectrum in the wavelength region redward of the $C_2$ (0,0) bandhead for the final C and N abundances because of uncertain molecular data. Therefore, the calculated atomic spectra  were fitted with the observed spectrum and the iterated abundance was accepted as an upper limit of the yttrium abundance. Spectral synthesis of the strontium  line at 4077.709 \AA\ confirms the low abundance of the first s$-$process peak (Figure~\ref{fig18}). 

\begin{figure}
\resizebox{\hsize}{!}{\includegraphics[angle = 0]{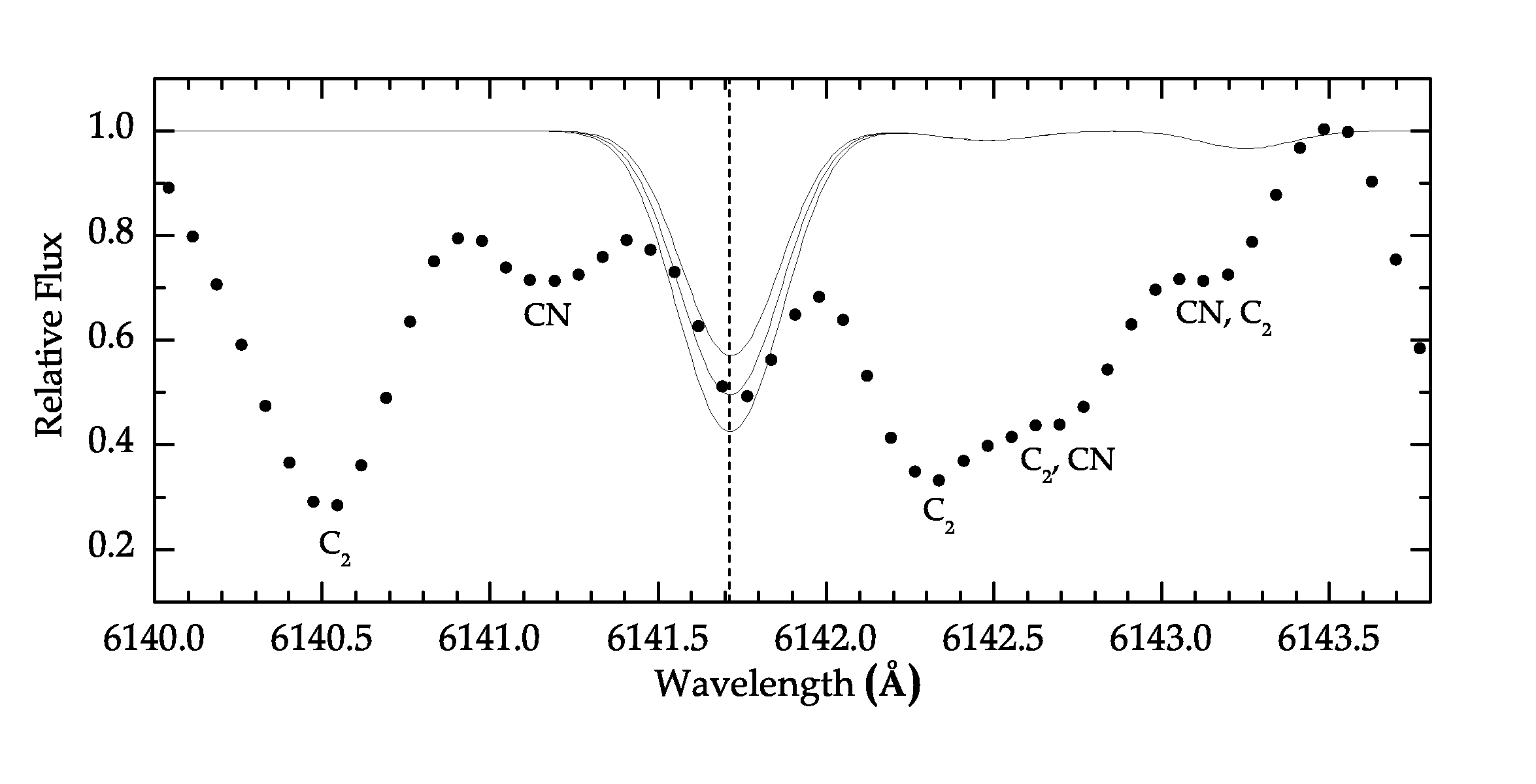}}
    \caption{The spectrum of HD\,112869 (filled circles), along with synthesized atomic spectra around Ba\,{\sc ii} lines at 6141.713 \AA. The synthesized spectra are shown for three barium abundances, $\ log\,\epsilon(Ba)$ = -0.7 $\pm$ 0.3 dex. } 
     \label{fig19}
\end{figure}

\begin{figure*}
\includegraphics[width=14cm]{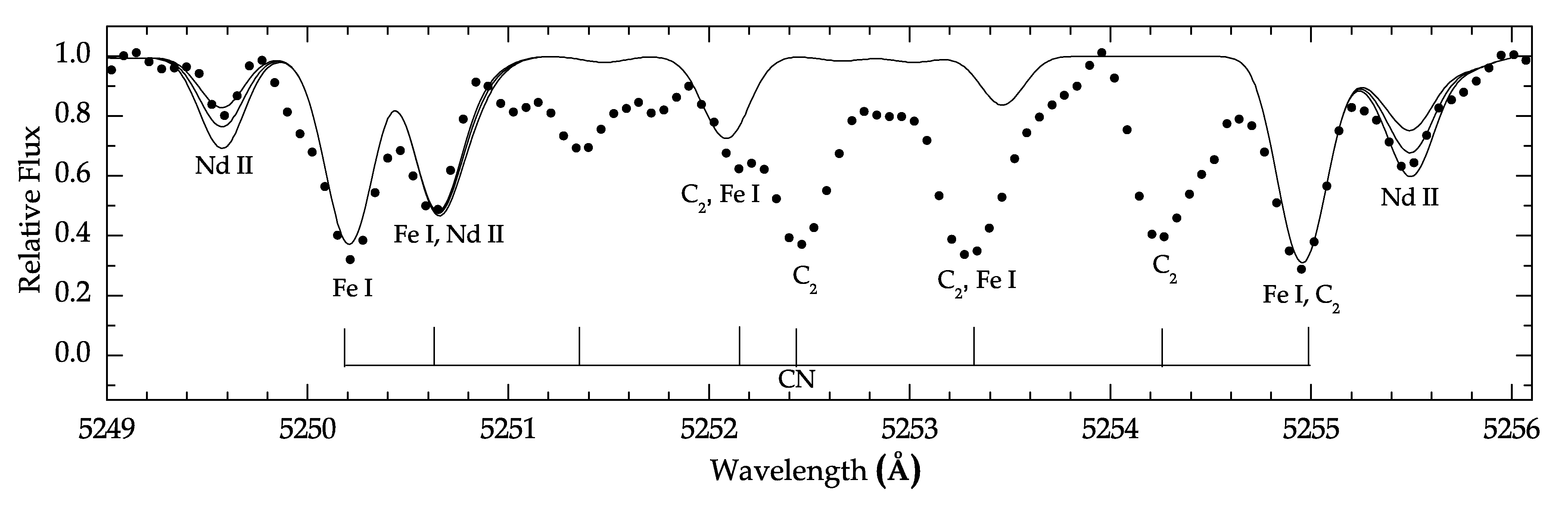}   
\caption{The spectrum of HD\,112869 (filled circles), along with synthesized  atomic spectra around Nd\,{\sc ii} lines at 5249.576 and 5255.502 \AA. The synthesized spectra are shown for three neodium abundances, $\ log\,\epsilon(Nd)$ = -0.2 $\pm$ 0.2 dex. }
    \label{fig20}
\end{figure*}

We selected a frequently employed for abundance analysis Ba\,{\sc ii} line at 6141.713 \AA\ but, we were not able to reproduce the neighboring $C_2$ and CN lines using the final atmospheric model because of the uncertain molecular data (Figure~\ref{fig19}). Again, the calculated atomic spectra  were fitted with the observed spectrum  and the iterated abundance was accepted as an upper limit of barium abundance, [Ba/Fe] $\la-$0.7 dex. NLTE corrections for the strontium and barium lines should be lower than 0.14 dex \citep{short}. The Ba\,{\sc ii} line at 4934.076 \AA\ confirms the absence of barium overabundance, [Ba/Fe] $\simeq -$0.1 dex.  Thus, two synthesized barium lines confirm a mild depletion of the heavy s$-$process element barium, [Ba/Fe] $< -$0.3 dex, on average. Spectral synthesis of two neodymium lines in the wavelength region redward of the $C_2$(0,0) bandhead shows the Nd enhancement (see Figure~\ref{fig20}). Synthesis of the La\,{\sc ii} line at 4086.709 and Sm\,{\sc ii} line at 4220.660 \AA\ confirms a similar overabundance. Thus, at least some of the elements of the second s$-$peak appear to be enhanced in the atmosphere of HD\,112868.  However, because of possibly weak unknown molecular blends we decided to be conservative and provide the upper abundance limits for the calculated neutron capture elements (Table~\ref{abundances}). A spectrum of very high resolution is needed to update the calculated abundances and to examine the third s$-$process peak which should be most enhanced according the theory of nucleosynthesis of metal-poor stars \citep{bisterzo10}.  

\begin{figure}
\resizebox{\hsize}{!}{\includegraphics{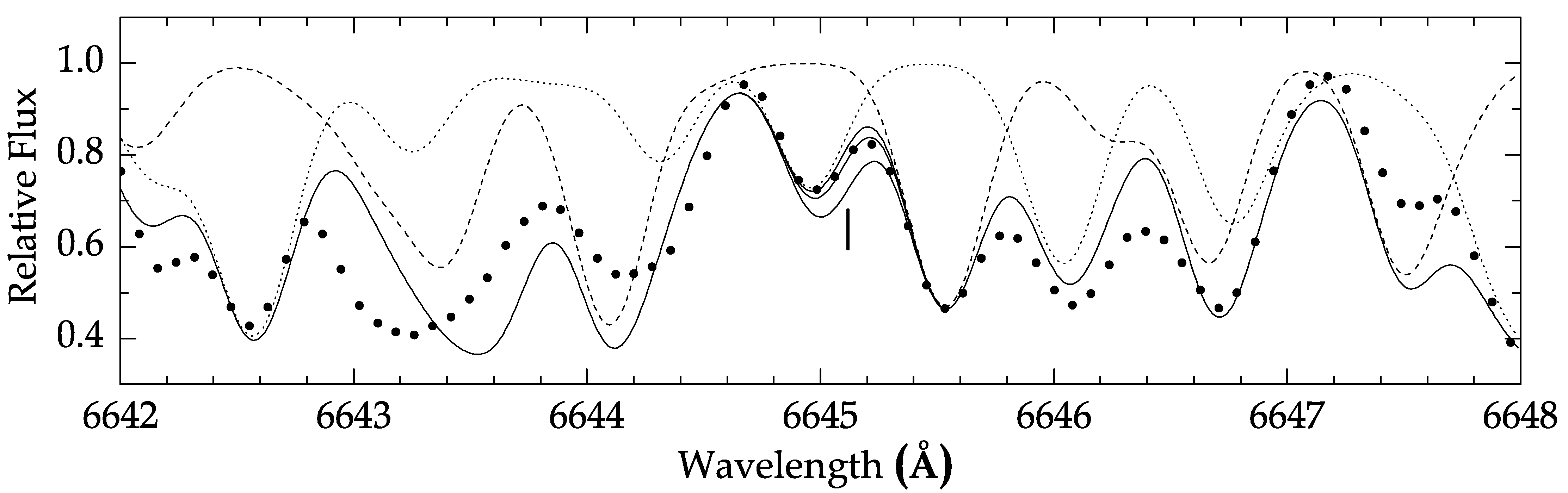}}
\caption{The spectrum of HD\,112869 (filled circles), along with synthesized spectra around the position of Eu\,{\sc ii} line at 6645.127 \AA\ marked by vertical tick. The latter are calculated for three europium abundances, $\ log\,\epsilon(Eu)$ = $-$0.5,  $-$1.0, and $-$1.5 dex (solid lines). The synthesized $C_2$ (dashed line) and CN (dotted line) spectra are given to illustrate the iterated molecular contribution.}
    \label{fig21}
\end{figure}

Our efforts to calculate the abundance of the r$-$process element europium was plagued by strong molecular blends and uncertain molecular data around the selected lines. The Eu\,{\sc ii} line at 6645.127 \AA\ was selected to be the best candidate for abundance analysis. However, we concluded during our calculations that the feature identified to be the europium line is blueshifted by about 6 km $s^{-1}$ relative to its rest wavelength. The main contributor to the feature at 6645.0  \AA, measured  in the spectrum of HD\,112869, was identified to be the CN line at 6644.961 \AA, and the Eu\,{\sc ii} line seems to be only a weak blend on the red wing of the CN line. Only a crude estimation of an upper limit of the europium abundance was possible because of uncertain oscillator strengths for the CN  line at 6644.961 \AA\ and for the neighboring $C_2$ lines at 6645.509 and 6645.611 \AA. We modified the oscillator strengths of these three molecular lines to fit the observed profiles and then estimated the contribution of europium.  A strong enhancement of r$-$process was denied for the atmosphere of HD\,112869, [Eu/Fe] $\la$ 0.8 dex (see Figure~\ref{fig21}).

\subsubsection{Comparison with the published abundances}

A comprehensive calculation of the atmospheric abundance for HD\,112869 from a high-resolution photographic spectrum was provided by \citet{kipper},  which resulted in the abundances for 26 elements. Unfortunatelly, his results are illustrated only in two narrow wavelength ranges around $C_2$ bandheads and the list of employed atomic  lines is absent. The iron abundance calculated by \citet{kipper} is lower that found in this paper and a difference of $\thicksim$0.6 dex seems to be mainly because of NLTE effects. \citet{kipper} calculated the iron abundance using neutral lines which are supposedly suffered from NLTE effects. \citet{aoki} studied the effects of metallicity for HD\,112869 using SED and the best fit with the observed spectral energy distribution for this star was achieved  using the model  with [Fe/H] $\eqsim -$2.0 (see Fig. 4 in \citet{aoki}. 

The carbon to oxygen ratio calculated by \citet{kipper} is typical of N-type carbon stars, C/O =  1.07, however, the adopted oxygen abundance is extremely high, $\log \epsilon$ (O) = 8.8 (see, also, comment in \citet{aoki}). \citet{aoki}  obtained the  C/O ratio in the range from 3.4 to 20 for six different atmospheric models with the calculated C/O value of 6.3 for the model with similar atmospheric parameters (see Table 1) and a the more reliable oxygen abundance, [O/Fe] = +0.5 dex. Thus, the C/O ratio calculated by us  is in line with that calculated by \citet{aoki}. The spectroscopic estimates of  oxygen abundance for HD\,112869 are still absent. Large discrepancy between the calculated nitrogen abundance ($\sim$2.3 dex) and those by \citet{kipper} follows partly from the difference in the adopted carbon abundance. Nobody has detected molecular isotopic lines in the spectrum of HD\,112869. Spectral synthesis of the selected wavelength regions near the position of isotopic $C_2$ and CN lines provides  a high lower limit for the carbon isotopic ratio of $^{12}C/^{13}C$  --  from 50 \citep{kipper} to 5000 \citep{tsuji, aoki} for six different models. Thus, our calculations are in line with the high $^{12}C/^{13}C$  ratio calculated by \citet{aoki}. 

\citet{kipper} found a large enhancement of neutron capture elements relative to the extremely low iron abundance ([Fe/H] = --2.9)  in the atmosphere of HD\,112869, [n/Fe] $\simeq$ +2.5 dex, on average, for 13 elements, which is in conflict with the present results. The reason of large discrepancy is not clear, however, \citet{kipper} has noted that because of heavy blending of lines the calculated abundances could be only the upper limits. Thus, the large overabundance of neutron-capture elements estimated by \citet{kipper} seems to be related to neglected molecular blending. Notice that we identified only a few relatively unblended lines of neutron capture elements in  the wavelength regions where molecular contamination is lower. Notice, too, that an overestimation of s$-$process abundances calculated  by \citet{kipper96} was detected  for the comparison star HD\,25408 (see Table~\ref{basic_dat}).

\section{Discussion and conclusions}

The radial-velocity monitoring of HD\,212869 shows variations with  a peak to peak amplitude of about 10~km s$^{-1}$ and a dominating period of about 114.9 days.  The variations are semiregular, with some changes from cycle to cycle and a significant scatter of velocities near the velocity extremums, which exceeds the standard error of measurements. Similar velocity variations and scatter are observed for two protoplanetary-nebulae -- HD\,235858 and BD+42$\degr$\,4388 \citep{hrivnak}. The velocity variations of HD\,112869 are accompanied by light and color variations. However, the light and color curves are shifted in phase relative to the velocity curve. The reason for the velocity, light, and color variations  is obviously the pulsations in the atmosphere of HD\,112869. The character of pulsations is similar to that observed for  evolved stars. The dominating period agrees well with the fundamental mode calculated for luminous ($\log L/L_{\sun} \simeq$ 3.2) and low mass (M = 0.8$M_{\sun}$) stars in the Large Magellanic Cloud (see Figure 25 in \citet{marigo}). 

The metallicity calculated for HD\,212869 is higher than that estimated prior to using high-resolution spectra. The iron abundance is found to be [Fe/H] = $-2.3 \pm$0.2 dex on the basis of ionised lines which are almost free from NLTE effects. The carbon abundance was found to be high in the atmosphere of HD\,112869, $\ log\,\epsilon(C)$ = 8.3 $\pm$ 0.1 dex. The nitrogen abundance is $\ log\,\epsilon(N)$ = 6.55 $\pm$ 0.2 dex. With the obtained  abundances [C/Fe] = +2.2 dex and [C/N] = +1.15, HD\,112869 occupies the region of CEMP-s stars on the plots   [C/Fe] vs.  [Fe/H]  and [C/N] vs.  [Fe/H]  (see Figures 5 and 6 in \citet{campbell}. However, the s--process elements Sr, Y, and Ba are not enhanced significantly, thus confirming by definition \citep{beers2005} the CEMP-no status of HD\,112869. However, the Nd abundance seems to be enhanced relative to iron and a similar overabundance was recognized for lanthanum and samarium. From inspection of the Eu\,{\sc ii} line at 6645.127 \AA, the upper limit was set for the r$-$process element europium, [Eu/Fe] $\la$ +0.8 dex. According to calculations carried out by \citet{bisterzo}, the abundances of three s-peaks are strongly dependent on the choice of the $^{13}C$-pocket as well as on the initial mass and the metallicity. [ls/Fe], [hs/Fe] and [Pb/Fe] do not follow a linear behavior with decreasing metallicity and can cover a large range of values. For the low mass models the enhancement of the first s-process peak is low or absent. A spectrum of very high-resolution  is needed to estimate the abundances for large number of the second s$-$process peaks and to recognize an overabundance of the third s$-$process peak. With the adopted oxgen abundance, [O/Fe] = +0.8 dex, the carbon to oxygen ratio was found to be very high for HD\,112869, C/O $\simeq$ 12.6. The isotopic lines of $C_2$ and CN are too week to be detected in the crowded spectrum and the lower limit of isotopic ratio was found to be extremely high, $^{12}C/^{13}C >$ 1500.   A large isotopic ratio is not typical of CEMP stars. However, for low-mass AGB stars the CN processing is not expected after the second dredge-up and a total amount of $^{12}C$ dredged-up during the AGB phase leads to a high [C/N] ratio and a high $^{12}C/^{13}C$ ratio observed for HD\,112869.  On the contrary,  intermediate-mass AGB stars with a hot bottom burning (HBB) should have low both [C/N] and $^{12}C/^{13}C$ ratios. 

Mass transfer in a binary system or evolution of a single star on the AGB  have been proposed to explain the abundance peculiarities observed for HD\,112869 \citep{tsuji, kipper, aoki}. The semiregular character of radial velocity variations accompanied by color variations rejects the binary status of HD\,112869. \citet{tsuji} discussed the final stages of evolution on the AGB with a small envelope mass to interpret the high C/O and $^{12}C/^{13}C$ ratios observed for HD\,112869.  A carbon to oxygen ratio above 1,  a high luminosity and a pulsational unstability are typical features of cool evolved AGB and post-AGB stars. HD\,112869, with the luminosity of $\log (L/L_{\sun}) \simeq 3.2$ and the effective temperature of $\log (T_{eff}) \simeq 3.6$,  lies on the AGB of  metal-poor (Z = 0.0001) low mass stars  according the evolutionary tracks  calculated by \citet{bertelli08}. During thermal pulses the photospheric C/O ratio can exceed 10. The polarization observed in the visual band confirms the presence of circumstellar matter around HD\,112869 \citep{goswami}. Thus, according to the current data, HD\,112869 seems to be a single metal-poor low mass TP-AGB star.

\acknowledgments
We acknowledge support from the Research Council of Lithuania under the grant No. MIP-85/2012. The European Union  FP7-PEOPLE-2010-IRSES program  is acknowledged for funding exchange visits in the framework of the project POSTAGBinGALAXIES (grant agreement No. 269193). LZ thanks Drs. T.Kipper, M.R.Schmidt, U.G.J\o rgensen, and A.Barzdis for assistance during the first iterations of abundance analysis  and valuable discussions.  This research has made use of the Simbad database operated at CDS (Strasbourg, France) and the VALD database operated at Uppsala University, the Institute of Astronomy RAS in Moscow, and the University of Vienna.


\begin{thebibliography}{}
\bibitem[Abia et al.(2002)] {abia02} Abia, C., Dominguez, I., Gallino, R., et al. 2002, ApJ, 579, 817  
\bibitem[Abia \& Isern(2000)] {abia00} Abia, C., \& Isern, J. 2000, ApJ, 536, 438 
\bibitem[Allen et al.(2012)] {allen} Allen, D.M., Ryan, S.G., Rossi, S., Beers, T.C., Tsangarides, S.A. 2012, A\&A, 548, A34
\bibitem[Aoki et al.(2007)] {aoki07} Aoki, W., Beers, T.C., Christlieb, N., et al. 2007, ApJ, 655, 492
\bibitem[Aoki et al.(2006)] {aoki06} Aoki, W., Frebel, A., Christlieb, N., et al. 2006, ApJ, 639, 897
\bibitem[Aoki et al.(2002)] {aoki02} Aoki, W., Norris, J.E., \&  Ando, H. 2002, PASJ, 54, 933
\bibitem[Aoki et al.(2004)] {aoki04} Aoki, W., Norris, J.E., \& Ryan, S.G. 2004, ApJ, 608, 971
\bibitem[Aoki \& Tsuji(1997)] {aoki} Aoki, W., \& Tsuji, T. 1997, A\&A, 317, 845 
\bibitem[Asplund(2005)] {asplund} Asplund, M. 2005, ARA\&A, 43, 481
\bibitem[Asplund et al.(2009)] {asplund2} Asplund, M., Grevesse, N., Sauval, A.J., \& Scott, P. 2009, ARA\&A, 47, 481 
\bibitem[Beers \& Christlieb(2005)] {beers2005} Beers, T.C., \& Christlieb, N. 2005, ARA\&A, 43, 531
\bibitem[Beers et al.(1992)] {beers92} Beers, T.C., Preston, G.W., \& Shectman, S.A. 1992, AJ, 103, 1987
\bibitem[Beers et al.(2007)] {beers07} Beers, T.C., Sivarani, T., Marsteller, B., et al. 2007, AJ, 133, 1193
\bibitem[Bergeat et al.(2001)] {bergeat01} Bergeat, J., Knapik, A., \& Rutily, B. 2001, A\&A, 369, 178
\bibitem[Bergeat et al.(2002)] {bergeat02} Bergeat, J., Knapik, A., \& Rutily, B. 2002, A\&A, 390, 967
\bibitem[Bertelli et al.(2008)] {bertelli08} Bertelli, G., Girardi, L., Marigo, P., \& Nasi, E. 2008, A\&A, 484, 815
\bibitem[Bisterzo et al.(2010)] {bisterzo10} Bisterzo, S., Gallino, R., Straniero, O., Cristallo, S., \& K\"appeler, F. 2010, MNRAS, 404, 1529
\bibitem[Bisterzo et al.(2011)] {bisterzo} Bisterzo, S., Gallino, R., Straniero, O., Cristallo, S., \& K\"appeler, F. 2011, MNRAS, 418, 284
\bibitem[Borysow et al.(1997)] {borysow} Borysow, U.G., J\o rgensen, U.G., \& Zheng, C.1997, A\&A, 324, 185
\bibitem[Campbell \& Lattanzio(2008)] {campbell} Campbell, S.W., \& Lattanzio, J.C. 2008, A\&A, 490, 769
\bibitem[Christlieb et al.(2001)] {christlieb01} Christlieb, N., Green, P.J., Wisotzki, L., \& Reimers, D. 2001. A\&A, 375, 366
\bibitem[Depagne et al.(2002)] {depagne02} Depagne, F., Hill, V., Spite, M., et al. 2002. A\&A, 390, 187
\bibitem[Eggen (1972)] {eggen} Eggen, O. 1972, ApJ, 174, 45  
\bibitem[Goswami \& Korinkuzhi(2013)]{goswami} Goswami, A., \& Korinkuzhi, D. 2013, A\&A, 549, A68
\bibitem[Griffin(1967)]{griffin} Griffin, R. 1967, ApJ, 148, 465
\bibitem[Harris et al.(2003)] {harris} Harris, G.J., Pavlenko, Y.V., Jones, H.R.A., \& Tennyson, J. 2003, MNRAS, 344, 1107
\bibitem[Harris et al.(2006)] {harris2} Harris, G.J., Tennyson, J., \& Kaminsky, B.M.  2006, MNRAS, 367, 400
\bibitem[Hirschi(2007)] {hirschi} Hirschi, R. 2007, A\&A, 461, 571
\bibitem[Hrivnak et al.(2013)] {hrivnak} Hrivnak, B.J., Lu, W., Sperauskas, J., et al. 2013, ApJ, 766, 116
\bibitem[J\o rgensen \& Larsson(1990)] {jorgensen} J\o rgensen, U.G., \& Larsson, M. 1990, A\&A, 371, 222
\bibitem[Keenan \& Morgan(1941)] {keenan} Keenan, P.C., \& Morgan, W.W. 1941, ApJ, 94, 501
\bibitem[Kennedy et al.(2011)] {kennedy} Kennedy, C.R., Thirupathi, S., Beers, T.C., et al. 2011, AJ, 141, 102
\bibitem[Kipper(1992)] {kipper} Kipper, T. 1992, BaltA, 1, 181  
\bibitem[Kipper et al.(1996)] {kipper96} Kipper, T., J\o rgensen, U.G., Klochkova, V.G., \& Panchuk, V.E. 1996, A\&A, 306, 489 
\bibitem[Kipper \& Kipper(1990)] {kipper90} Kipper, T., \& Kipper, M.A. 1990, SvAL, 16, 478  
\bibitem[Kornilov et al.(1991)]{WBVR91} Kornilov, V. G., Volkov, I. M., Zakharov, A. I., et al. 1991, Catalogue of WBVR magnitudes of bright stars of the northern sky, ed. V. G. Kornilov, Tr. Gos. Astron. Inst. im. Sternberga, 63, 400 p.
\bibitem[Kupka et al.(1999)] {kupka} Kupka, F., Piskunov, N., Ryabchikova, T.A., Stempels, H. C.; \& Weiss, W. W. 1999, A\&AS, 138, 119
\bibitem[Kurucz(1993)] {kuruczs} Kurucz, R.L., 1993, CD-ROM 1--22, Smithsonian Astrophysical Observatory 
\bibitem[Kurucz(2005)]{kurucz05} Kurucz, R.L. 2005, Mem. S.A.It. Suppl., 8, 14 
\bibitem[Landolt(1992)]{landolt92} Landolt, A. U. 1992, AJ, 104, 340
\bibitem[Lucatello(2006)]{lucatello06} Lucatello, S., Beers, T.C., \& Christlieb, N. 2006, ApJ, 652, L37
\bibitem[Lucatello et al.(2005)]{lucatello05} Lucatello, S., Gratton, R.G., Beers, T.C., \& Caretto, E. 2005, ApJ, 625, 833
\bibitem[Marcy et al.(1989)]{marcy} Marcy, G.W., \& Benitz, K.L. 1989. ApJ, 344, 441
\bibitem[Marigo \& Girardi(2007)]{marigo} Marigo, P., \& Girardi, L. 2007, A\&A, 469, 239
\bibitem[Mashonkina(2013)]{mashonkina} Mashonkina, L. 2013, A\&A, 550, A28 
\bibitem[Masseron et al.(2010)] {masseron10} Masseron, T., Johnson, J.A., Plez, B., et al. 2010,  A\&A, 509, A93 
\bibitem[Merle et al.(2011)] {merle} Merle, T., Th\'evenin, F., Pichon, B., \&  Bigot, L. 2011, MNRAS, 418, 863 
\bibitem[Meynet et al.(2006)]{meynet} Meynet, G., Ekstrom, S., \& Maeder, A. 2006, A\&A, 447, 623
\bibitem[Nidever et al.(2002)]{nidever} Nidever, D.L., Marcy, G.W., Butler, R.P., Fische, D.A., \& Vogt, S.S. 2002, ApJS, 141, 503
\bibitem[Norris et al.(2001)]{norris01} Norris, J.E., Ryan, S.G., \& Beers, T. 2001, ApJ, 561, 1034
\bibitem[Pavlenko(1997)] {pavlenko97} Pavlenko, Y.V. 1997, Ap\&SS, 253, 43 
\bibitem[Pavlenko(2003)] {pavlenko03} Pavlenko, Y.V.  2003, Astron. Zh., 80, 67
\bibitem[Pavlenko \& Zhukovskaya(2003)] {pz} Pavlenko, Y.V., \& Zhukovskaya, S.V. 2003, Kinem. Fiz. Neb. Tel, 19, 28 
\bibitem[Piskunov et al.(1995)] {VALD1} Piskunov, N. E., Kupka, F., Ryabchikova, T. A., Weiss, W. W., \& Jeffery, C. S. 1995, A\&AS, 112, 525 
\bibitem[Samus et al.(2009)]{samus} Samus, N.N., Durlevich, O.V., \& Kazarovets, E.V. 2009, General Catalogue of variable Stars, Vizier On-Line Data Catalog 
\bibitem[Shchukina et al.(2005)] {shchukina} Shchukina, N.G., Bueno, J.T., \& Asplund, M. 2005, ApJ, 618, 939
\bibitem[Short \& Hauschildt(2006)] {short} Short, C.I., \& Hauschildt, P.H. 2006, ApJ, 641, 494
\bibitem[Sivarani et al.(2006)] {sivarani} Sivarani, T., Beers, T.C., Bonifacio, P., et al. 2006, A\&A, 459, 125  
\bibitem[Sneen et al.(1976)] {sneen} Sneen, C., Johnson, H.R., \& Krupp, B.M. 1976, ApJ, 204, 281
\bibitem[Tanaka et al.(2007)] {tanaka} Tanaka, M., Letip, A., Nishimaki, Y., et al. 2007, PASJ, 59, 939
\bibitem[Tsuji et al.(1991)] {tsuji} Tsuji, T., Iye, M., Tomioka, K., Okada, T., \& Sato, H. 1991, A\&A, 252, L1
\bibitem[Udry et al.(1999)]{udry1} Udry, S., Mayor, M., Maurice, E., et al.  1999, in Precise Stellar Radial
   Velocities, eds. J.B. Hearnshaw \& C.D. Scarfe, APS Conf., 185, 383.
\bibitem[Umeda  \& Nomoto(2005)]{umeda} Umeda, H., \& Nomoto, K. 2005, ApJ, 619, 427
\bibitem[Upgren et al.(2002)]{upgren} Upgren, A., Sperauskas, J., \& Boyle, R. P. 2002, BaltA, 11, 91
 \bibitem[van Leeuwen(2007)] {vanLeeuwen} van Leeuwen, F.  2007, A\&A, 474, 653
\bibitem[Yoss \& Griffin(1997)] {yoss} Yoss, K.M., \& Griffin, R.F. 1997, JApA, 18, 161
\bibitem[Zamora et el.(2009)] {zamora} Zamora, O., Abia, C., Plez, B., Dominguez, I., \& Cristallo, S. 2009, A\&A, 508, 909  
\end{thebibliography}
\end{document}